\documentclass[preprint]{elsarticle}
\usepackage{lineno,hyperref}
\usepackage[utf8]{inputenc}
\journal{Journal of Information Sciences}

\usepackage{color}
\usepackage{multirow,longtable,booktabs}

\usepackage{graphicx}
\usepackage{booktabs}
\usepackage{enumitem}
 \setlist{itemsep=.2em}

\usepackage[format=plain,
  labelformat=simple,
  font={sf},
  indention=0cm,
  skip=0pt,
  labelsep=period,
  singlelinecheck=false,
  tableposition=top,
  figureposition=bottom,
]{caption}
\usepackage{subcaption}
\usepackage{comment}
\usepackage{float}
\usepackage{amsmath}
\usepackage{amsfonts}
\usepackage{amssymb}
\usepackage{cases}
\begin{document}

\begin{frontmatter}

%Tweets2Stance: Users stance detection exploiting Zero-Shot Learning Algorithms from Tweets
%: Exploiting Zero-Shot Learning Algorithms to Stance-detection from Tweets
\title{$Tweets2Stance$: Users stance detection exploiting Zero-Shot Learning Algorithms on Tweets}%\tnoteref{mytitlenote}
% oppure 'Exploiting Zero Shot Learning Algorithms for political stance detection using Tweets'?

\author[mymainaddress,mysecondaryaddress]{Margherita Gambini}%\fnref{myfootnote}
%\address{CNR - Istituto di Informatica e Telematica (IIT), Pisa, Italy}
\cortext[mycorrespondingauthor]{Corresponding author}
\ead{m.gambini@iit.cnr.it}
\author[mymainaddress]{Tiziano Fagni}
\ead{t.fagni@iit.cnr.it}
\author[mymainaddress]{Caterina Senette\corref{mycorrespondingauthor}}
\ead{c.senette@iit.cnr.it}
\author[mymainaddress]{Maurizio Tesconi}
\ead{m.tesconi@iit.cnr.it}



\address[mymainaddress]{Institute of Informatics and Telematics (IIT) - CNR, Via Giuseppe Moruzzi, 1 56124 Pisa – Italy}
\address[mysecondaryaddress]{Engineering School - University of Pisa,  Largo Lucio Lazzarino, 2, 56122 Pisa - Italy}

\begin{abstract}
In the last years there has been a growing attention towards predicting the political orientation of active social media users,
being this of great help to study political forecasts, opinion dynamics modeling and users polarization. Existing approaches, mainly targeting Twitter users, rely on content-based analysis or are based on a mixture of content, network and communication analysis. The recent research perspective exploits the fact that a user's political affinity mainly depends on his/her positions on major political and social issues, thus shifting the focus on detecting the stance of users through user-generated content shared on social networks.
The work herein described focuses on a completely unsupervised stance detection framework that predicts the user's stance (along five levels of agreement) about specific social-political statements by exploiting content-based analysis of its Twitter timeline. The ground-truth user's stance may come from Voting Advice Applications (VAAs), online tools that help citizens to identify their political leanings by comparing their political preferences with party political stances. 
Starting from the knowledge of the agreement level of six parties
on 20 different statements (VAA’s statements), the objective of the study is to
predict the stance of a Party $p$ in regard to each statement $s$ exploiting what
the Twitter Party account wrote on Twitter. To this end we propose $Tweets2Stance$ ($T2S$), a novel and totally unsupervised stance detector framework which relies on the zero-shot learning technique to quickly and accurately operate on non-labeled data. Interestingly, T2S can be applied to any social media user for any context of interest, not limited to the political one. Results obtained from multiple experiments (analysis, measurement, and evaluation) show that, although the general maximum F1 value is $0.4$, $T2S$ can correctly predict the stance with a general minimum MAE of $1.13$, which is a great achievement considering the task complexity.
%In the near future, we will investigate the use of this framework to predict the political leanings of users.
\end{abstract}

\end{frontmatter}

%\linenumbers

\section{Introduction}
%intercept consumption preferences, anticipate needs, foresee or manipulate different user leanings
%si potrebbe proprio partire con lo spiegare che la stance detection è utile per vari motivi (incluso political leaning), ma che le varie tecniche hanno limitazioni (data sparsity, data collection etc.). Quindi noi introduciamo un framework che indirizza tali limitazioni ed è di tipo user-level, target-specific. Poi diciamo che abbiamo preso spunto dalle VAA, e che quindi i nostri esperimenti si focalizzano sugli utenti = partiti politici
\label{sec:intro}
During the last years, there has been a growing attention towards social media as for what is explicitly shared among users (content, thoughts, and behavior), as well as about what is hidden and latent. 
%noticeably, what is not explicitly said represents a precious source of information to intercept consumption preferences, anticipate needs, foresee or manipulate different user leanings. 
Among this latent information, the user's stance, i.e. the expression of a user's point of view and perception toward a given statement \cite{biber1988adverbial}, is particularly interesting; in fact, stance detection on social media is an emerging opinion mining paradigm that well applies in different social and political contexts, and for which many researchers are working to propose solutions ranging from natural language processing, web science, and social computing \cite{aldayel2021stance, dias2016inf,igarashi2016tohoku,augenstein2016stance,mohammad2016semeval,hamidian2019rumor, darwish2017improved,darwish2020unsupervised}. 

Many works \cite{aldayel2021stance} dealt with the stance detection at the user-level; however, to the best of our knowledge, a completely unsupervised technique exploiting user's textual content only has never been explored.
%for political and social issues \cite{aldayel2021stance}: 
%knowing active users' political stances is beneficial for the political discourse analysis, and allow politicians, political observers, and strategists to adequately set up their campaigns, investigate the performance of a Party or a candidate, and bring forward their vulnerabilities to mitigate them and improve election campaigns; moreover, the social media customers' stance about products is vital to plan the next business move; besides, the fake news detection and claim validation research fields also benefit from the stance-detection [citare qualcosa].
%However, the task of detecting the social media user's stance toward a given statement has got several issues, the most difficult being the data collection and data sparsity: the data sparsity problem refers to the fact that real-world data concerning users’ stances is sparse, but this problem is now less relevant in the era of BigData; instead, data collection requires to gather user's stances, and this task can be trivial when users are directly polled about that, or it could be a very challenging problem when stances need to be detected among user-generated content shared on social media. 
Hence, the work herein described focuses on the design and implementation of a stance detection framework named $Tweets2Stance$ ($T2S$), able to detect the stance of a Twitter account using its timeline. 
The idea for this framework stems out from observing how Voting Advice Applications (VAAs) work.
%we predict the agreement/disagreement with a statement, expression of a political Party, exploiting content-based analysis from tweets and ground-truth data coming from Voting Advice Applications (VAAs).
Voting Advice Applications, originally developed in the 1980s as paper-and-pencil civic education questionnaires \cite{cedroni2010voting}, are online tools that aid citizens, mainly before elections, to identify their political leaning by comparing their policy preferences with the political stances of parties or candidates running for office. VAAs are widespread in many countries and have a crucial role in online election campaigns worldwide. Basically, the user marks its positions on a range of policy statements. The application compares the individual's answers to the positions of each Party or candidate and generates a rank-ordered list or a graph indicating which Party or candidate is located closest to the user's policy preferences. One of the crucial elements of the VAAs is the questionnaire: the selection of the questions, their balance among the political poles, and their phrasing have an impact both on the way in which users respond, as well as on the overall users‘ engagement on the poll itself. For these reasons, the VAA's issued statements should cover the spectrum of the most important topics of an election campaign and adequately show crucial differences among all the competitors on political scenario for which the VAA is designed\cite{louwerse2014design}. This careful definition of the questionnaire, i.e. taking into high consideration the main topics under discussion at a certain time, suggested us the possibility of using the official position of Italian parties about specific political statements (during a certain political election period) as the ground-truth to determine that stance from the timeline of the Twitter Party accounts in a completely unsupervised way \footnote{the Italian Parties' official positions about 20 political statements were kindly provided by the Observatory on Political Parties and Representation \cite{political_observatory} based on the VAA \textit{NavigatoreElettorale} for the European Elections 2019 \cite{Navigatore_Elettorale_2019}}; notice that only tweets written during the pre-election period are considered.
%it in combination with the textual content that the users post on their Twitter accounts during the pre-election period. 

\paragraph{Objectives} Starting from the knowledge of the agreement level of six parties on 20 different statements (VAA's statements), the objective of the study is to predict the stance of a Party $p$ in regard to each statement $s$ exploiting what the Twitter Party account wrote on Twitter. According to a recent literature review \cite{aldayel2021stance}, our work can be classified as a stance detection task at the user level, since it relies on the textual content of Twitter posts (timelines of parties’ Twitter accounts) with a further specialization as target-specific, the most common form of stance-detection on social media; here, the target identifies a topic (whose stance is predicted), but differently from previous works in the literature our classification model is built for different topics. 
Focusing on content-based approaches, some limitations that emerge in existing solutions are: (i) intrinsic difficulty of processing natural language; (ii) need for large corpora of manually-annotated tweets and language-specific resources; (iii) supervised approaches, which lack generalizability and suffer from the limited availability of comprehensive and reliable ground-truth datasets \cite{cohen2013classifying}. 

To overcome these drawbacks, we focus on the newest pre-trained language models based on the Transformer architecture \cite{vaswani2017attention} which currently allow to achieve the best effectiveness on practically all important supervised learning NLU tasks \cite{dong2019unified}. Specifically, we rely on an unsupervised technique called zero-shot learning (ZSL) \cite{chang-2008-dataless_class} to exploit pre-trained models as generic classifiers to infer the stance of a party towards a given topic of interest (see subsection \ref{subsec:ZSC} for more details).
Moreover, differently from state-of-the-art methods, our goal is to come up with a fine-grained stance-detector solution working along five classes that could be generalized to various spheres, not just the political one.
\paragraph{Contribution}The main contributions of the work can be summarized as follows:
\begin{itemize}
    \item To the best of our knowledge, our $Tweets2Stance$ ($T2S$) framework represents an innovative method to stance-detection task; it uses a bottom-up approach based on stance-detection centered on textual content posted on the Party’s Twitter accounts and a set of VAA statements on which we know the official position taken by the Party.
    \item As detailed in the Related Work section \ref{sec:related_work}, our contribution is innovative since we offer a relatively simple and totally unsupervised context-based method where the contents are not individual texts but the users' timelines. Our framework can be applied to the timelines of other social media as well.
    \item The proposed $T2S$ technique is organized as a generic framework which can be easily customized to be adapted to specific social media user requirements. In this work, the tested social media users are six Italian parties, since we had their official stance about 20 political statements \cite{political_observatory, Navigatore_Elettorale_2019}.
    \item 
    %The described approach applies well to the political scenario, since it is the first step of a pipeline that allows to predict the political orientation (having the ground truth to measure its performance);
    The described approach applies well to the political scenario, since it is the first logical step of an inference pipeline useful to predict the users' political orientation;
    nevertheless, this technique can also be generalized to
    other contexts, such as extremism/radicalization to infer whether a user
    has become radicalized on certain themes (immigration, vaccines, etc.). 
    \item The proposed approach is easily adaptable to different languages, even in the absence of pretrained ZSL models specific to the language of interest. In this study we focus on the Italian language and compare various ZSL models to deal with it.
    %\item: vantaggi: tecnica è sempre la stessa ma può aumentare l'accuratezza mano a mano che escono modelli di ZSL migliori
    \end{itemize}
\paragraph{Road-map}
The remainder of the paper is organized as follows: Section \ref{sec:related_work} discusses the related work, while Section \ref{sec:task_def} clearly defines the prediction task on which our framework will work. Section \ref{sec:data_collection_pre_processing} describes the preliminaries of the study detailing the political scenario under analysis and the data collection and cleaning procedures. Then, Section \ref{sec:methodology} details our framework for the prediction task. Afterwards, Sections \ref{sec:experimental_setup} and \ref{sec:results_and_discussion} describe the experimental setup and the results achieved. Finally, Section \ref{sec:conclusion} ends the paper by drawing conclusions and future works.

\section{Related Work}
\label{sec:related_work}
Stance-detection is an emerging opinion mining paradigm that well applies in several social and political scenarios. The state of the art resumed in a highly valuable recent survey \cite{aldayel2021stance} highlights the importance of categorization, since stance-detection can be classified according to the target (single, multi-related, or claim-based) or to the task type (detection or prediction).
Moreover, in order to better narrow the related work, we need to distinguish between works at the user level and works at the statement level. 
At the statement level \cite{murakami2010support,walker2012your}, whose objective is to predict the stance described in a piece of text, previous research works are mainly based on Natural Language Processing (NLP) methods and classification tasks with three classes (support/against/none).
Instead, at the user level, the objective is to predict the stance of a user toward a given topic and generally, prediction solutions incorporate different users’ attributes along with the text of their posts. The work of Aldayel and Magdy \cite{aldayel2019your} examined how various users' online features such as their posts, the network they interact with or follow, the websites they visit, and the content they like can reveal their stance towards different topics. In this case, rather than focusing on the detection accuracy, the authors' main focus was to understand how stance could be revealed testing multiple sets of features on a stance benchmark dataset of different topics.
Lynn et al. \cite{lynn2019tweet} investigated the predictive power of user-level features alone compared to document-level features for document-level tasks (it means predicting the stance of a tweet without the tweet). To this end, they present some NLP tasks that might be better primarily cast as user-level, supporting the importance on integrating user features into predictive systems.
\subsection {Stance detection target-specific}
According to the mentioned literature Review \cite{aldayel2021stance}, our work could be classified as a stance detection task at the user level with a further specialization as target-specific, the most common form of stance-detection on social media. Here the target identifies a topic (whose stance is predicted) and generally a separated classification model is built and trained for each topic. Approaches described in literature \cite{ gottipati2013predicting, dias2016inf,igarashi2016tohoku} are based on the text incorporated in a post together with multiples post/users attributes; in those cases, task classification mainly used two classes, support and against. Some examples are detailed in the following.
In the work of Gottipati et al. \cite{gottipati2013predicting}, the authors proposed a collaborative filtering approach to solve the data sparsity problem of users stances on ideological topics, and apply clustering method to group the users with the same party.  The experiments showed that using ideological stances with Probabilistic Matrix Factorization (PMF) technique achieved a high accuracy of 88.9\% at 22.9\% data sparsity rate and 80.5\% at 70\% data sparsity rate on users’ party prediction task. The proposed approach was applied to bipolar stance-detection (two positions per target). 
In the work of Dias and Becker \cite{dias2016inf}, a weakly supervised solution for detecting stance in tweets was described; it used n-grams representing opinion targets, common terms denoting stance (e.g. hashtags) and sentiment strategies to build a large corpus representing the input of a supervised prediction algorithm. 
In the work of Igarashi et al. \cite{igarashi2016tohoku}, the authors compared feature-based and Neural Network based approaches on the supervised stance classification task for tweets. They achieved the best results with Convolutional Neural Network (CNN) based approach in the cross validation on the training data, and the worst result on the test data due to the CNN overfitting; in contrast, the feature-based approach was more robust, leveraging the external knowledge.
Other target-specific strategies in literature were conducted at the statement level, such as those proposed by Augenstein et al. \cite{augenstein2016stance},Mohammad et al. \cite{mohammad2016semeval}, Sardar Hamidian et al. \cite{hamidian2019rumor}, Yingjie Li et al. \cite{yin-etal-2019-benchmarking}, and Umme Aymun Siddiqua \cite{siddiqua2019tweet}. 
The strategy presented in \cite{augenstein2016stance} was to use conditional LSTM encoding (Long short-terms memory networks), which built a target-dependent representation of the tweet and it outperformed the technique of encoding the tweet and the target independently. Performance was further improved by augmenting conditional model with bidirectional encoding. The approach was conducted at the statement level through unsupervised methods, and classification was made along three positions (favour, against, neither).
In \cite{mohammad2016semeval} the authors reported results obtained involving different teams in a stance-detection shared task. The task was to infer a three-level stance of tweets using automatic natural language systems: in favor of the given target, against the given target, or neither inference is likely. To this end, they subdivided the task into a supervised (task A) and an unsupervised (task B) sub-task, receiving 19 and 9 submissions from teams respectively. The highest classification F-score obtained was 67.82 for Task A and 56.28 for Task B.
In the work of Sardar Hamidian et al. \cite{hamidian2019rumor}, the authors performed a rumor stance classification experimenting different levels of pre-processing through supervised classificators; they exploited a new set of meta linguistic and pragmatic features, and performed the experiments with and without pre-processing on the textual content. Their approach did not seem to benefit from pre-processing.
In the work of Yingjie Li et al. \cite{yin2019benchmarking}, the authors designed a framework which incorporates a target-specific attention mechanism and sentiment classification, outperforming the state of the art of deep learning methods.
The goal of Siddiqua et al.'s work \cite{siddiqua2019tweet} was to determine the stance of a tweet according to three positions. To this end, a particular neural network model was employed: it adopted the strengths of two LSTM variants coupled with an attention mechanism that amplified the contribution of important elements in the final representation. As a result, the method went over the current state-of-the-art deep learning based methods both for single and multi-target benchmark datasets.

As mentioned above, target-specific approaches could consider single or multiple targets. Usually, the concept of multi-target classification has been used to analyse the relation between two political candidates by using domain knowledge about these targets to improve the classification performance. In that case, the same model can be applied to different targets on the hypothesis that the same piece of text that contains the stance in favour to a target, it also implicitly contains the stance against the other \cite{sobhani2017dataset, darwish2017improved, lai2018stance}. Obviously, those approaches need multi-target stance-detection datasets. In the political context examined in our study, each statement is representative of a specific target, thus we must face a multi-target classification task in the broadest sense, that is not limited to a target and its opposite. Previously cited methods well apply on one target at a time, since they operate on datasets containing texts that have already been classified as relevant to the topic (whose stance is predicted); consequently, these strategies need pre-filtered data containing additional elements such as particular hashtags or keywords. On the contrary, our method does not need to work with selected texts, because it includes a data filtering step which guarantees in input only relevant data. Moreover, our approach works well in the multi-target scenario without differentiating the classification model for each target thanks to the advantage of operating with a generic zero-shot learning classifier. 
\subsection {Machine Learning approaches}
Considering features to model stance-detection tasks and machine learning (ML) algorithms applied, we focus on linguistic cues, as they are the element we are interested in. Here, the literature reports a further subdivision on (i) linguistic features, i.e. the most used approach that reveals the stance based on text-linguistic features \cite{ghosh2019stance, mohammad2016semeval}, and (ii) users' vocabulary (stance-detection is based on the user's choice of vocabulary) \cite{darwish2020unsupervised, dong2019unified}.
Since textual cues could refer both to textual features, sentiment, and semantics, we limit our attention to textual features. 
Given the above, and focusing on machine learning approaches used to model and train classifiers employed on all the so far mentioned works, the most are based on \textit{supervised} ML techniques \cite{gottipati2013predicting,igarashi2016tohoku,lai2018stance,walker2012your,zhang2020enhancing}.
Some works attempted to enrich datasets entities applying \textit{unconstrained supervised} methods such as transfer learning, weak-supervision, and distant supervision methods for stance detection \cite{augenstein2016stance,dias2016inf}.
Others innovative approaches are those that propose \textit{unsupervised learning} strategies \cite{darwish2020unsupervised,joshi2016political,fagni2022} exploiting clustering techniques and embeddings representations of users’ tweet\cite{rashed2020embeddings}.
\bigskip\\
The major limitations that emerge from all the aforementioned studies are represented by:
\begin{itemize}
    \item For the approaches that exploit textual and non-textual information in stance-detection, data collection and analysis are highly time-consuming, especially in the case of network-based solutions;
    \item The required data are often not available or very difficult to retrieve, mainly due to increasing data protection policies adopted by Social Media over the years;
    \item The majority of the aforementioned detection models have only two (maximum three) stance classes;   
\item Almost all the mentioned techniques require supervised or semi-supervised prediction models whose limits, well known in literature \cite{cohen2013classifying}, include, among others,  the need for large amounts of data and the poor generalizability, being strongly dependent on the dataset used as training-set. Among unsupervised attempts it is worth mentioning the work of Darwish et al. \cite{darwish2020unsupervised} close related to our study, that used dimensional reduction to project users onto a low-dimensional space, followed by a clustering step, which allowed to find core users that are representative of the different stances. Their proposed solution relied on unsupervised methods not requiring prior labeling of users, neither a domain- or topic-level knowledge, but needing a manually labeled cluster (low effort). Comparing it with our work, the main emerging differences are: (i) Darwish's method only analyses users who produce the most of the posts on a given topic, instead ours approach considers any type of user, even those who have very few (1 or 2) significant tweets (non-vocal users); (ii) The Darwish method works on 2 classes, our method detects five classes; (iii) The Darwish method provides a top-down solution to stance-detection starting from a group analysis that includes the users who discuss the topic the most and trying to separate them into two classes (for, against). The Darwish method is not suitable for making individual analysis if there is no context information deriving from other users, i.e. it cannot analyse a single user and understand if he is in favour or against a certain topic. Our method, on the other hand, is a bottom-up approach that analyses the information of the target user without the need to have access to network or context information from other users (e.g. retweets); (iv) Moreover, in the Darwish method the stance detection must be done on pre-filtered data relevant to the topic of interest. Our method, having set the sentence and the associated topic, can work on any type of content, appropriately self-filtering relevant data for the topic.
\end{itemize}
For all these reasons, the recent challenge to user-level and target-specific stance detection is to move towards unsupervised systems exploiting textual content only. To this aim, a ZSL technique exploiting advanced pre-trained Natural Language Inference (NLI) models \cite{ghosh2019stance, yin2019benchmarking} can be a viable solution (see Section \ref{subsec:ZSC} for further details), as our stance-detection framework proved. All things considered, our work may open the way to a novel approach to stance-detection both at the user and target-specific level (as detailed in the subsequent sections); to the best of our knowledge, this particular stance-detection task is not yet described in the closely related literature, hence representing the first of further contributions from/to the community of researchers working on this topic.

%For all these reasons, the recent challenge is to move towards unsupervised systems to predict continuous and non-discrete affinity values which could be best achieved when the ground-truth values were available for training and evaluating the system. In line with this objective are the pre-trained NLI models that are gaining more and more attention \cite{ghosh2019stance, yin2019benchmarking}. Grabbing the novelties introduced by such recent research and going beyond the state of the art, our work could open the way to a novel approach to stance-detection both at the user and target-specific level (as detailed in the subsequent sections); to the best of our knowledge, this particular stance-detection task is not yet described in the closely related literature, hence representing the first of further contributions from/to the community of researchers working on this topic.

%For all these reasons, the recent challenge to user-level and target-specific stance detection is to move towards unsupervised systems. To this aim, the advanced pre-trained NLI models \cite{ghosh2019stance, yin2019benchmarking} can be a viable solution, as our stance-detection framework will show.

\section{Task Definition}
\label{sec:task_def}
\begin{figure}[ht]
    \centering
    \includegraphics[scale=0.8]{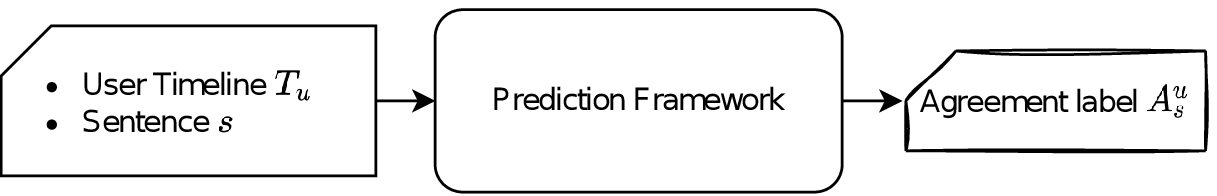}
    \caption{The task is predicting the agreement level $A^u_s$ of Twitter User $u$ in regard to sentence $s$ exploiting the User's textual content timeline $TL_u$.}
    \label{fig:Fig1}
\end{figure}

%To the best of our knowledge, it is the first work that try to infer the political stance of a Party \textit{p} by measuring the agreement/disagreement level between the textual content posted on the Party's Twitter account and a set of VAA statements on which we know the official position taken by the Party.
The task (Fig. \ref{fig:Fig1}) is to predict the stance $A^u_s$ of a Social Media User $u$ with respect to a social-political statement (or sentence) $s$ making use of the User's textual content timeline on the considered social media (e.g., the Twitter timeline).

The stance $A^u_s$ is a five-valued categorical label, as shown in Table \ref{tab:agreement_disagreement_labels}.

\begin{table}[!htbp]
	\sf\centering
	\caption{The agreement/disagreement levels}
	\begin{tabular}{lc}
		\toprule
		\multirow{2}{*}{Categorical} & Integer\\
		& Mapping\\
		\midrule
		completely agree & 5\\
		agree & 4\\
		neither disagree, nor agree & 3\\
		disagree & 2\\
		completely disagree & 1\\
		\bottomrule
	\end{tabular}
	\label{tab:agreement_disagreement_labels}
\end{table}

The desired ground-truth is the label $G^u_s$, that is the known agreement/ disagreement level of User $u$ in regard to sentence $s$. Remind that the ground-truth is only used to evaluate our proposed $Tweets2Stance$ framework and find its optimal parameters; no training step ever occurs. In this work, we assume that users are the Twitter accounts of six Italian Parties, as the following section will detail.

%In this work we assume that users are Political Accounts

%As explained in Section \ref{sec:data_collection_pre_processing}, the Twitter timelines of the six major Italian Parties were used.

\section{Data collection and Pre-processing}
\label{sec:data_collection_pre_processing}

The political scenario under analysis refers to the European and Municipal elections in Italy on 26th May 2019, when Italian citizen were called for the election of the Italian representatives to the European Parliament. The number of Members of the European Parliament (751 deputies in total) for each country is approximately proportional to the population. In 2019, Italy had to elect 76 deputies. Contextually, Italian voters had also to participate to municipal election of mayors, municipal and district councillors (in about 3800 Italian municipalities), with planned run-off on 9th June 2019.
In that context, we focused our attention on the six major parties in Italy: three center-right parties including Forza Italia (FI), Fratelli d’Italia (FDI), and Lega, two left-wing parties including Partito Democratico (PD) and +Europa (+Eu)\footnote{+Europa was recently born in 2018 and it is characterized for a pro-European and liberal orientation.}, and the Movimento 5 Stelle (M5S) representing a sort of third pole at that time. The Italian parliament included other minor parties, especially on the left- wing, representing less than 5\% of the Italian population each. We did not consider these parties in the current study.
As previously said, we started from the assumption that knowing the parties' answers on the VAA's statements, it is possible to predict the stance of a Party $p$ in regard to each statement $s$ exploiting what the Party wrote on Twitter. The definition of the 20 statements (Table \ref{tab:topic_definition_eng} and Table \ref{tab:topic_definition_ita}) that express the political positions of the six referenced parties towards selected themes under discussion in Italy and in Europe in 2019, was entrusted to a group of political experts \cite{political_observatory, Navigatore_Elettorale_2019} who provided us with the ground-truth $G^p_s$ for each Party $p$ and statement $s$ on which the current work is based.

At first, we collected timelines  of the official Twitter account of each party using the official Twitter API\footnote{\url{https://developer.twitter.com/en/docs}}.
Considering the speed with which the political discussion nowadays takes place especially on social media, the observation period was adequately chosen in order to maximize the number of tweets avoiding noise and off-topics content. Furthermore, to intercept any valuable information or discussion trends overtime we have extended the analysis considering four temporal ranges and built the associated datasets\footnote{ The four raw datasets can be found at \url{https://github.com/marghe943/Tweets2Stance_dataset}} as described in Table \ref{tab:datasets}.

\begin{table}[!htbp]
	\sf\centering
	\caption{The four studied datasets with the total number of tweets before the pre-processing step. $D_j$ contains $j$ months of tweets.}
	\begin{tabular}{p{0.19\columnwidth}p{0.2\columnwidth}p{0.2\columnwidth}p{0.2\columnwidth}p{0.2\columnwidth}}
		\toprule
		 & $D_3$ & $D_4$ & $D_5$ & $D_7$\\
		\midrule
		Period & [2019-03-01, 2019-05-25] & [2019-02-01, 2019-05-25] & [2019-01-01, 2019-05-25] & [2018-11-01, 2019-05-25]\\
		\#tweets & 20'266 & 25'979 & 34'736 & 44'370 \\
		\bottomrule
	\end{tabular}
	\label{tab:datasets}
\end{table}

As a preliminary step, since the text collected from tweets contains a lot of noise and irrelevant information, 
%and cannot be directly used as input for machine learning models, 
we pre-processed the tweets in order to remove anything which don’t have predicting significance, that is: 
\begin{itemize}
    \item URLs (uniform resource locator), $++$, $\&gt$, $\&lt$, \textbackslash$n$ were replaced with an empty string, 
    \item we removed the "$RT @user: $" prefix of retweets, 
    \item we removed the mentions at the beginning of a reply tweet,
    \item the hashtags were replaced with empty string, being difficult to analyse by the Zero-Shot Classifier component (see Section \ref{subsec:ZSC}),
    \item the emojis were replaced with an empty string, being difficult to analyse by the Zero-Shot Classifier component (see Section \ref{subsec:ZSC}),
    \item heading whitespaces were replaced with an empty string,
    \item more than two subsequent whitespaces were replaced with a single whitespace,
    \item finally, we removed tweets with $\{1,2,3\}$ words and empty tweets.

\end{itemize}

Lastly, since we wanted to test our prediction approach on English tweets as well (as explained in Section \ref{subsec:ZSC}), we further translated the Italian tweets using the \textit{google\_trans\_new}\footnote{\url{https://pypi.org/project/google-trans-new/}} Python package.

\section{Methodology}
\label{sec:methodology}
\begin{figure}[ht]
    \centering
    \includegraphics[scale=0.6]{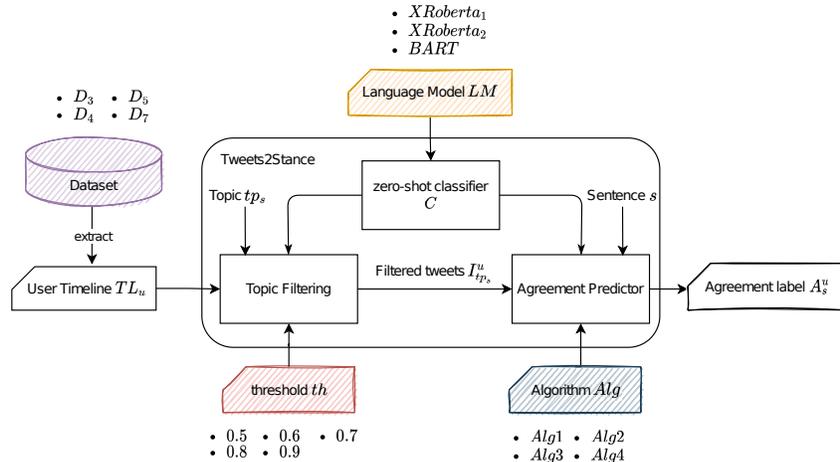}
    \caption{Our Tweets2Stance framework to compute the agreement level $A^u_s$ of User $u$ in regard to sentence $s$. The inputs are the Twitter timeline $TL_u$ extracted from a certain dataset $D_i$, the sentence $s$, the topic $tp$ associated to $s$, a language model $LM$, a threshold $th$ and an algorithm $Alg$. The highlighted components are the parameters that we'll vary during our experiments, as explained in Section \ref{sec:experimental_setup}. 
    %The output is the triplet $(p,s,A^p_s)$.
    }
    \label{fig:Fig2}
\end{figure}

%[discutere in maniera generale. Poi, ci sono 3 sottomoduli su cui agire. Una sottosezione per ognuno dei 3 sottomoduli; usare una lettera per formalismo]
This section presents the proposed $Tweets2Stance$ (T2S) framework (Fig. \ref{fig:Fig2}) to predict the stance $A^u_s$ of a Twitter User $u$ in regard to a sentence $s$, exploiting its Twitter timeline $TL_u = [tw_1,...,tw_n]$. A User might either not talk about a specific political argument (here expressed with sentence \(s\)), or debate on an issue not risen by our pre-defined set of statements. 
%or debate on an issue not risen by the other considered sentences
For these reasons, our framework executes a preliminary $Topic$ $Filtering$ step, exploiting a zero-shot classifier $C$ to get only those tweets talking about the topic $tp$ of the sentence $s$. Finally, the $Agreement$ $Predictor$ uses the in-topic tweets $I^u_{tp_s}$ and the same zero-shot classifier $C$ to predict the agreement/disagreement level $A^u_s$. 

As Fig. \ref{fig:Fig2} illustrates, the parameters of the $T2S$ framework to tune are four:
\begin{enumerate}
    \item the language model $LM$ to use for the zero-shot classification in both the $Topic$ $Filtering$ and the $Agreement$ $Predictor$ modules,
    \item the dataset $D_i$ from which extracting the timeline $TL_u$,
    \item the algorithm $Alg$ to use in the $Agreement$ $Predictor$ module,
    \item the threshold $th$ to get the in-topic tweets $I^u_{tp_s}$ in the $Topic$ $Filtering$ module.
\end{enumerate}
Through the experiments described in Section \ref{sec:experimental_setup}, we performed a grid search over the four parameters to get the optimal setup.
%for this prediction task

The following subsections will discuss each of the three main modules of our framework: the zero-shot classifier $C$ to comprehend what a tweet is about, and the $Topic$ $Filtering$ and $Agreement$ $Predictor$ modules. Hereinafter, we'll detail the proposed framework into a specific political scenario where the Twitter accounts of interest are those of the six Italian parties reported in Section \ref{sec:data_collection_pre_processing}, hence the User $u$ becomes the Party $p$.
%together with the selection of the language model $LM$ to experiment on
%then it defines the $T2S$ method, followed by the definition of the topic $tp$ for each sentence and the description of the proposed algorithms $Alg$s.

%Then, it defines the topic $tp$ for each sentence, followed by the discussion on the choice of the zero-shot classification (ZSC) framework as part of the solution and on the selected language models $LM$ to experiment on. Finally, the proposed algorithms $Alg$s used for the $Agreement$ step of our $T2S$ method are described.

%presents the proposed $T2S$ method in Fig. \ref{fig:Fig2} to predict the stance $A^p_s$ of a Party $p$ in regard to a sentence $s$ exploiting its Twitter timeline $TL_p$. Then, it defines the topic $tp$ for each sentence, followed by the discussion on the choice of the zero-shot classification (ZSC) framework as part of the solution and on the selected language models $LM$ to experiment on. Finally, the proposed algorithms $Alg$s used for the $Agreement$ step of our $T2S$ method are described.

\subsection{Zero-shot Classification} %Zero-shot Learning technique
\begin{figure}[ht]
    \centering
    \includegraphics[scale=0.8]{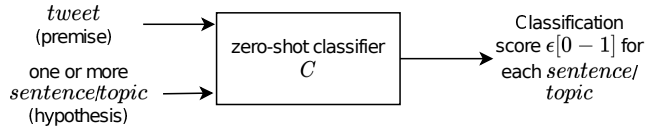}
    \caption{A zero-shot classifier receives in input a tweet (premise) and one or more sentence/topic (hypothesis). The output is the classification score for each couple (tweet, sentence or topic).}
    \label{fig:Fig3}
\end{figure}

\label{subsec:ZSC}
To comprehend the textual semantic of a tweet, we relied on the advanced Natural Language Processing (NLP) methods based on recent and powerful neural language models \cite{qiu2020pre}, since they have shown extremely competitive performance on many popular Natural Language Understanding (NLU) tasks \cite{radford2019gpt2, devlin-etal-2019-bert, brown2020gpt3}. Particularly, we exploited the zero-shot learning technique first proposed by \cite{chang-2008-dataless_class} to use language models pre-trained on the Natural Language Inference (NLI) task \cite{yin-etal-2019-benchmarking} as generic classifiers able to quickly and accurately operate on non-labelled data. This specific classification method is called \textit{zero-shot-classification} (Fig. \ref{fig:Fig3}).

%is a specific type of ZSL, that is teaching a classifier on one set of labels, and then using it on a set of labels that the classifier may have never seen before.
%%Notice that a label may also be a sentence or a sequence of keywords. 
%Since our Twitter dataset is not labelled in any way with respect to the agreement/disagreement level on the 20 sentences, the ZSC framework seems the optimal tool to find out what the user is talking about in his/her Twitter timeline. %%choice to compare tweets and sentences. 
%Subsection \ref{subsec: Two-Step Method} will detail how the ZSC was used.
%% https://nlp.town/blog/zero-shot-classification/#:~:text=Zero%2Dshot%20classifiers%20predict%20the,hundreds%20of%20labeled%20training%20items.
%%https://www.analyticsvidhya.com/blog/2020/12/understanding-text-classification-in-nlp-with-movie-review-example-example/
%https://discuss.huggingface.co/t/new-pipeline-for-zero-shot-text-classification/681/63
%https://jaketae.github.io/study/zero-shot-classification/
%https://arxiv.org/abs/1712.05972
%%Introduzione allo zero shot learning e il posto che ha nel nostro modello, va spiegato bene %perché abbiamo scelto questo e va descritto il problema linguistico

%\subsubsection{Language Model Choice for the Zero-shot Classifier}
%\label{subsubsec:zsc_model_choice}
The HuggingFace Transformer library \cite{wolf-etal-2020-transformers} provides ready-to-use state-of-the-art neural language models (the ones based on the Transformer architecture \cite{vaswani2017attention}) along with an easy framework to work on most of the NLU tasks. %it is constantly updated with the newest language model architectures, and it supplies a unique and easy framework to work on most of the NLU tasks. 
For this reason, the HuggingFace library was chosen to perform zero-shot classification.
%%% 
Many neural %Transformer-based 
language models were pre-trained solely on English-language texts (e.g., GPT2 \cite{radford2019gpt2} and BERT \cite{devlin-etal-2019-bert}). Unfortunately, our Twitter datasets contain tweets written in Italian. There are currently three ways of addressing this English-centricity: 
\begin{enumerate}
    \item \label{item_eng_centricity:a} translating the documents in English and using the English models, or
    \item relying on models pre-trained on a single non-English Language (e.g., the Italian AlBERTo \cite{polignano_2019_alberto}, the Portuguese's PTT5 \cite{carmo_2020_ptt5}, the Dutch BERTje \cite{devries2019bertje}), or
    \item \label{item_eng_centricity:c} using multi-lingual models that have been pre-trained on a mixture of many languages (XLM \cite{conneau_2019_xlm}, XLM-Roberta \cite{conneau-etal-2020-unsupervised}, mBART \cite{liu_2020_mbart}, mT5 \cite{xue-etal-2021-mt5}). %The core is to share a vocabulary created through Byte Pair Encoding (BPE), which greatly improves the alignment of embedding spaces across languages (\cite{conneau_2019_xlm}. 
\end{enumerate}
%The HuggingFace zero-classification framework works only 
Since the zero-shot classification stems from the NLI task, only those language models pre-trained or already fine-tuned on one or more NLI tasks can be used, which is not the case for the Italian Transformer-based language models (AlBERto \cite{polignano_2019_alberto}, BERTino \cite{Muffo2020BERTinoAI}). 
Therefore, we chose to follow the \ref{item_eng_centricity:a}) and \ref{item_eng_centricity:c}) solutions, in particular adopting these 3 configurations:
\begin{description}
    \item[BART] The BART-large model \cite{lewis-etal-2020-bart} fine-tuned on the MultiNLI dataset \cite{williams_2018_mnli_dataset} ($BART\_Large_{MNLI}$, \cite{bart_large_mnli}) was used with tweets translated in English. We exploited the library \textit{google\_trans\_new}\footnote{\url{https://pypi.org/project/google-trans-new/}}, a wrapper for Google Translate service, to perform automatic translation of tweets from Italian to English language.
    \item[XRoberta\_1] The XLM-Roberta-Large model \cite{conneau-etal-2020-unsupervised} fine-tuned on the XNLI dataset \cite{conneau-etal-2018-xnli} was employed as multi-lingual model over Italian tweets ($XLM\_Roberta$ $\_Large_{XNLI}$, \cite{joeddav_xlm-roberta-large-xnli}). Notice that the XNLI dataset doesn't include Italian, but the authors assure that since the base model (XLM-Roberta-Large) was pre-trained on 100 different languages (Italian included), the fine-tuned model has shown good performances in NLI task even with other languages available only in base model.
    \item[XRoberta\_2] We evaluated the \(XLM\_Roberta\allowbreak\_Large_{IT\_MNLI}\) \cite{Jiva_xlm-roberta-large-it-mnli} on Italian tweets: this model takes XLM-Roberta-Large and fine-tunes it on an Italian translation of a subset of the MNLI dataset (85\% of the train set).
\end{description}

\subsection{Topic Filtering}
\label{subsec:topic_filtering}

%Let's consider a sentence $s$ and a certain dataset $D-Nmonths$ from which extracting the Twitter timeline $TL_p = [t_1,...,t_n]$ of Party $p$. The goal of our proposed $T2S^p_s$ method is to predict the agreement/disagreement value \(A^p_s\in\{1,2,3,4,5\}\).

%A Party might either not talk about a specific political argument (here expressed with sentence \(s\)), or debate on an issue not risen by the considered VAA sentences. For these reasons, our $T2S^p_s$ method executes a preliminary 
The $Topic$ $Filtering$ module outputs the in-topic tweets $I^p_{tp_s}$ to be fed to the $Agreement$ $Predictor$ module. To this aim, we had to define a topic for each of the considered statements: Table \ref{tab:topic_definition_eng} and Table \ref{tab:topic_definition_ita} show the English and Italian topic definitions for the $20$ statements that were used for all experiments; as Section \ref{sec:experimental_setup} will describe, we experimented both with Italian tweets and the same ones translated in English by varying the language model $LM$ used for C.
%The idea is to define a topic for each of the considered statements, so that the 

\begin{longtable}[!ht]{p{0.05\columnwidth} p{0.6\columnwidth} p{0.4\columnwidth}}
    \caption{Defined topic for each of the \(20\) sentence (English version).}\\
    \toprule
    nr. & Sentence & Topic \\
    \midrule
    \endfirsthead
    \multicolumn{3}{c}%
    {\tablename\ \thetable\ -- \textit{Continued from previous page}} \\
    \toprule
    nr. & Sentence & Topic \\
    \midrule
    \endhead
    \midrule \multicolumn{3}{r}{\textit{Continued on next page}} \\
    \endfoot
    \bottomrule
    \endlastfoot
    1 & overall, being EU members is a disadvantage & European Union disadvantages \\
    2 & Italy should exit the euro & exit the euro \\
    3 & a common European army should exist & common European army \\
    4 & multinational corporations like Google and YouTube should pay copyright and taxes according to the rules of each European country & taxes for multinationals compliant with rules of each European country \\
    5 & european economic integration has driven too far: member states should regain greater autonomy & members economic autonomy in the European Union\\
    6 & the European Union should reform its immigration policy: Italy should receive more support from other Member States & immigration handling in EU \\
    7 & Italy should strengthen economic cooperation with China & economic cooperation between China and Italy\\
    8 & the recreational use of cannabis should be legal & recreational use of cannabis \\
    9 & Islam is a threat to Italy's values & Islam threatens Italian values \\
    10 & women must be guaranteed autonomy of choice over abortion & autonomy of choice over abortion \\
    11 & any form of self-defense within the private property should be legitimate & self-defense in your own home \\
    12 & the activities of the judiciary must be independent of political pressures & independence of the judiciary from politics \\
    13 & children who were born in Italy to foreign parents should receive Italian citizenship automatically & Italian citizenship for children born in Italy to foreign parents \\
    14 & wealth should be redistributed from wealthier citizens to poorer citizens & redistribution of wealth to the poorest \\
    15 & companies should be able to dismiss employees more easily & ease of dismissal of employees \\
    16 & healthcare should be more open to private operators & healthcare open to private operators \\
    17 & protecting the environment is more important than the economic growth & environmental protection versus economic growth \\
    18 & cutting public spending is a good way to solve the economic crisis & public spending cuts as a solution for the economic crisis \\
    19 & income support for the poorest sections of the population is positive for the Italian economy & improve the economy by helping low-income segments of population \\
    20 & the introduction of a single tax rate on income ("Flat Tax") would be beneficial for the Italian economy & taxes with a single rate without progressive taxation \\
    \label{tab:topic_definition_eng}
\end{longtable}

\begin{longtable}[!ht]{p{0.05\columnwidth} p{0.6\columnwidth} p{0.4\columnwidth}}
    \caption{Defined topic for each of the \(20\) sentence (Italian version).}\\
    \hline
    nr. & Sentence & Topic \\
    \hline
    \endfirsthead
    \multicolumn{3}{c}%
    {\tablename\ \thetable\ -- \textit{Continued from previous page}} \\
    \hline
    nr. & Sentence & Topic \\
    \hline
    \endhead
    \hline \multicolumn{3}{r}{\textit{Continued on next page}} \\
    \endfoot
    \hline
    \endlastfoot
    1 & nel complesso, essere membri dell'UE è uno svantaggio & svantaggi dell'Unione Europea \\
    2 & l'Italia dovrebbe uscire dall'Euro & uscire dall'euro \\
    3 & dovrebbe esistere un esercito comune europeo & esercito europeo comune \\
    4 & le multinazionali come Google e Youtube dovrebbero pagare i diritti d'autore e le tasse secondo le regole di ciascun paese europeo & tasse per le multinazionali in relazione alle regole di ciascun Paese Europeo \\
    5 & l'integrazione economica europea si è spinta troppo oltre: gli Stati membri dovrebbero riguadagnare maggiore autonomia & autonomia economica dei membri dell'Unione Europea\\
    6 & l'Unione Europea dovrebbe riformare la propria politica dell'immigrazione: l'Italia dovrebbe ricevere più supporto dagli altri Stati membri & gestione dell'immigrazione nell'Unione Europea \\
    7 & l'Italia dovrebbe intensificare le sue relazioni economiche con la Cina & relazioni economiche dell'Italia con la Cina\\
    8 & l'uso ricreativo della cannabis dovrebbe essere legale & uso ricreativo della cannabis \\
    9 & l'Islam è una minaccia per i valori dell'Italia & minaccia dell'Islam nei confronti dei valori italiani \\
    10 & alle donne deve essere garantita autonomia di scelta sull'aborto & autonomia di scelta sull'aborto \\
    11 & ogni forma di auto-difesa all'interno della proprietà privata dovrebbe essere legittima & legittima difesa nella propria abitazione con armi \\
    12 & le attività della magistratura devono essere indipendenti dalle pressioni della politica & indipendenza della magistratura dalla politica \\
    13 & i bambini, nati in Italia da cittadini stranieri, dovrebbero ricevere la cittadinanza italiana automaticamente & cittadinanza italiana per bambini nati in Italia da famiglie straniere \\
    14 & la ricchezza dovrebbe essere redistribuita dai cittadini più abbienti ai cittadini più poveri & redistribuzione della ricchezza verso i piu poveri \\
    15 & le imprese dovrebbe poter licenziare i dipendenti più facilmente & possibilita delle imprese di licenziare facilmente i propri dipendenti \\
    16 & la Sanità dovrebbe essere più aperta agli operatori privati & apertura della Sanità ad operatori privati \\
    17 & proteggere l'ambiente è più importante della crescita economica & importanza della protezione dell'ambiente \\
    18 & tagliare la spesa pubblica è un buon modo per risolvere la crisi economica & tagli alla spesa pubblica come soluzione per la crisi economica \\
    19 & il sostegno al reddito alle fasce più povere della popolazione è positivo per l'economia italiana & migliorare l'economia aiutando le fasce a basso reddito \\
    20 & l'introduzione di una aliquota unica sui redditi ("flat tax") sarebbe di beneficio all'economia italiana & conseguenze della flat tax per l'economia italiana \\
    \label{tab:topic_definition_ita}
\end{longtable}

In detail, the $Topic$ $Filtering$ module (Fig. \ref{fig:Fig2}) receives in input the Twitter Timeline $TL_p$ of Party $p$ and the topic $tp_s$ related to sentence $s$; by exploiting the zero-shot classifier $C$, the in-topic tweets $I^p_{tp_s}$ and the relative topic scores $T^p_{tp_s}$ are retrieved
\begin{equation}
    I^p_{tp_s} = \{tw_1,...,tw_m |C(tw_i, tp_s) >= th\}
\end{equation}
\begin{equation}
    T^p_{tp_s} = \{C(tw_i, tp_s) | tw_i \in I^p_{tp_s}\}
\end{equation}
%where $C$ is fed with a tweet $tw_i\in TL_p$ (the premise) and $tp_s$ (the hypothesis); 
The C's default hypothesis reformulation "\textit{This text is about \(<tp_s>\).}" was used. $C(tw_i, tp_s)\in[0,1]$ indicates how much tweet $tw_i$ refers to topic $tp_s$. $th$ is the filtering threshold value that we varied during the experiments defined in Section \ref{subsec:experiments_detail}.

\subsection{Agreement Predictor}
Afterwards, the $Agreement$ $Predictor$ module (Fig. \ref{fig:Fig2}) computes the final five-valued label \(A^p_s\) through an algorithm $Alg(T^p_{tp_s},S^p_s)$, defining
\begin{equation}
    S^p_s = \{C(tw_i, s) | tw_i \in I^p_{tp_s}\}
\end{equation}
as the C scores of tweets $I^p_{tp_s}$ with respect to sentence $s$; each score suggests how much tweet \(tw_i\) refers to and agrees with the sentence \(s\). In this case, the C hypothesis (the sentence $s$) was not reformulated, but used as it is.

%\subsubsection{Proposed Algorithms}
%\label{subsubsec:proposed algorithms}
%$STEP2$ requires an algorithm $Alg$ to compute $A^p_s$.

%Being $al\in \{1,2,3,4,5\}$ an agreement/disagreement level, where $3$ indicates "\textit{neither disagree, nor agree}", for each algorithm
%\begin{equation}
%    A^p_s = 
%    \begin{cases}
%      3 & \text{if $\mid I^p_{tp_s} \mid = 0$}\\
%      al & \text{otherwise}
%    \end{cases}   
%\end{equation}
%This is justified by the heuristic that if a user has never talked about a certain $s$, it is likely that he/she neither disagrees, nor agrees with it.

Each employed algorithm $Alg$ exploits one of the following mapping functions:
\begin{equation}
    M1(s) = 
    \begin{cases}
      1 & \text{if $s\in [0,0.2)$}\\
      2 & \text{if $s\in [0.2,0.4)$}\\
      3 & \text{if $s\in [0.4,0.6)$}\\
      4 & \text{if $s\in [0.6,0.8)$}\\
      5 & \text{if $s\in [0.8,1]$}\\
    \end{cases}
\end{equation}
or
\begin{equation}
    M2(s) = 
    \begin{cases}
      1 & \text{if $s\in [0,0.25)$}\\
      2 & \text{if $s\in [0.25,0.5)$}\\
      3 & \text{if $s\in [0.5,0.75)$}\\
      4 & \text{if $s\in [0.75,1]$}\\
    \end{cases}
\end{equation}
where the $M1(s)$'s integer values from $1$ to $5$ are related to the five agreement/disagreement labels defined in Table \ref{tab:agreement_disagreement_labels}, while $M2(s)$'s integer values from $1$ to $4$ are related to an intermediate agreement/disagreement Likert scale: $\{1,2\}$ have got the same meaning as in Table \ref{tab:agreement_disagreement_labels}, then $3$ indicates $agree$ and $4$ stands for $completely$ $agree$; the reason behind this intermediate mapping will be explained in $Alg4$ (subsection \ref{item:alg_4}). $s\in [0,1]$ refers to a score value both for $M1$ and $M2$.
%where $nl$ indicates whether the neutral label 3 must be considered or not.
%\textit{Step 2} (\ref{item:step2}) requires an algorithm $Alg$ to compute the final label $A\_p_p\_s_s$ for Party $p_p$ in regard to sentence $s_s$. For each algorithm, if there are no in-topic $tweets\_p_p\_s_s$ for a certain topic $tp_{s_s}$, the final label for sentence $s_s$ is set to \(3\) (\textit{neither disagree, nor agree}). This is justified by the heuristic that if a user has never talked about a certain $s_s$, it is likely that he/she neither disagrees, nor agrees with it. Also, each algorithm exploits the C classification \(score\in[0-1]\) to map either a final score or the $sc\_t_i\_s_s$ of each tweet to the agreement/disagreement $A\_level\in \{1,2,3,4,5\}$; this mapping occurs with equally-spaced intervals, since unequal spaces [finire]

The proposed algorithms ordered by complexity are the followings:

\begin{description}
    \item [Algorithm 1] \label{item:alg_1}\textbf{[Alg1]} The label $A^p_s$ is computed as % using $T^p_{tp_s}$ as values and $S^p_s$ as weights
    \begin{equation}
        A^p_s = 
        \begin{cases}
            M1(\frac{\sum_{i=1}^{|I^p_{tp_s}|}s_i\cdot t_i}{\sum_{i=1}^{|I^p_{tp_s}|}s_i}) & \text{if $\mid I^p_{tp_s} \mid \neq 0$}\\
            3 & \text{otherwise}
        \end{cases}
        %weighted\_avg(IT^p_{tp_s}, IT^p_s)
        %\varnothing
    \end{equation}
    where $s_i \in S^p_{tp_s}$ and $t_i \in T^p_{tp_s}$. 
    %Then, $A^p_s$ is the mapping of $w\_avg$ over the integer $al\in \{1,2,3,4,5\}$
    %\begin{equation}
    %    A^p_s = 
    %    \begin{cases}
    %        M1(w\_avg) & \text{if $w\_avg \neq \varnothing$}\\
    %        3 & \text{otherwise}
    %    \end{cases}
    %\end{equation}
    %It computes the weighted average over the sentence scores $sc\_t\_s_s$ of the in-topic tweets $tweets\_p_p\_s_s$ using the topic scores $sc\_t\_tp_{s_s}$ as weights. Finally, the average score is mapped to a five-valued integer $\{1,2,3,4,5\}$ representing the final agreement/disagreement level $A\_p_p\_s_s$ of Party $p_p$ in regard to sentence $s_s$.
    \item [Algorithm 2] \textbf{[Alg2]} First, it maps each tweet $tw_i\in I^p_{tp_s}$ into the label $l_i\in \{1,2,3,4,5\}$ using its sentence score $s_i\in S^p_s$
    \begin{equation}
        l_i = M1(s_i)
        \label{eq:al_i_assignment}
    \end{equation}
    then, $A^p_s$ is
    \begin{equation}
        A^p_s = 
        \begin{cases}
            \Bigl\lfloor \frac{\sum^{|I^p_{tp_s}|}_{i=1}l_i}{|I^p_{tp_s}|} \Bigr\rceil & \text{if $\mid I^p_{tp_s} \mid \neq 0$}\\
            3 & \text{otherwise}
        \end{cases}
        \label{eq:avg_a_i_labels}
    \end{equation}
    %It maps each in-topic tweet $t_i\in tweets\_p_p\_s_s$ into a five-valued \(label\_t_i\in\{1,2,3,4,5\}\) using its sentence score $sc\_t_i\_s_s$; $label\_t_i$ represents how much tweet $t_i$ refers to sentence $s_s$. Then, $A\_p_p\_s_s$ is the nearest integer to the average of all labels $label\_t_i$.
    
    %For each sentence, the in-topic tweets (tweets that passed the \textit{threshold\_value}) are collected. Then, the following steps are executed:
    %\begin{enumerate}[a.]
    %    \item assign the final label \(\{1,2,3,4,5\}\) to each in-topic tweet using its sentence score:
    %        \begin{itemize}
    %            \item label \(1\) (completely disagree) \(if\) \(sentence\_score < 0.2\)
    %            \item label \(2\) (disagree) \(if\) \(0.2<= sentence\_score < 0.4\)
    %            \item label \(3\) (neither disagree, nor agree) \(if\) \(0.4<= sentence\_score < 0.6\)
    %            \item label \(4\) (agree) \(if\) \(0.6<= sentence\_score < 0.8\)
    %            \item label \(5\) (completely agree) \(if\) \(0.8<= sentence\_score <=1\)
    %        \end{itemize}
    %    \item count how many tweets have been labelled (the number of voters) for each of the five final labels. If there are more than two final labels having the same and maximum number of voters, then the final label for the current sentence is the standard mean on in-topic tweets' final label (rounded by the nearest integer); otherwise, the final label for the current sentence is the label having the majority of voters.
    %\end{enumerate}
    
    The step of assigning \(l_i\) to each tweet $tw_i\in I^p_{tp_s}$ (Eq. \ref{eq:al_i_assignment}), hopefully returns a more fair $A^p_s$. 
    %In fact, it may resemble the macro/micro-metric issue: by aggregating the sentence scores (e.g. by weighted average as in $Alg1$), the macro aggregation is applied; instead, by aggregating the contribution of each tweet ($al_i$) using the standard mean, the micro aggregation is applied. In a multi-class classification setup, micro-metric is preferable if it is suspected that there may be class imbalance; in fact, the values $al_i$ are not balanced with respect to the current sentence \(s\): likely, if a Party $p$ agrees with a sentence, there will be lot of tweets in agreement with it (many $al_i=4$ or $al_i=5$) and a few (errors) or no tweets in disagreement (few labels $al_i=1$, or $al_i=2$, or $al_i=3$), and vice-versa. 
    In fact, the tweet normalization may help in aggregating the contribution of each tweet ($l_i$) using the standard mean, which means applying the macro aggregation. In a multi-class classification setup, macro-metric aggregation is preferable if it is suspected that there may be class imbalance; in fact, the values $l_i$ are not balanced with respect to the current sentence \(s\): likely, if a Party $p$ agrees with a sentence, there will be lot of tweets in agreement with it (many $l_i=4$ or $l_i=5$) and a few (errors) or no tweets in disagreement (few labels $l_i=1$, or $l_i=2$, or $l_i=3$), and vice-versa. 
    \item [Algorithm 3] \textbf{[Alg3]} Like $Alg2$, but slightly modifying how $A^p_s$ is computed (Eq. \ref{eq:avg_a_i_labels}). Let's further define $V_{l}$ as the number of voters for the integer label $l\in \{1,2,3,4,5\}$
    \begin{equation}
        V_l = |\{l_i : l_i = l\}^{|I^p_{tp_s}|}_{i=1}|
        \label{eq:num_voters}
    \end{equation}
    where $l_i$ are the labels computed from Eq. \ref{eq:al_i_assignment}. Let's define $v = max(V_l)$, then
    \begin{subnumcases}{A^p_s = }
        l  & \text{if $|\{l:V_l=v\}| = 1$}\label{eq:sub_case_majority_voting}\\
        \Bigl\lfloor \frac{\sum^{|I^p_{tp_s}|}_{i=1}l_i}{|I^p_{tp_s}|} \Bigr\rceil & \text{if $|\{l:V_l=v\}| > 1$}\label{eq:sub_case_mean}\\
        3 & \text{otherwise}
    \end{subnumcases}
    %Like \textit{Algorithm 2}, but introducing the majority voting as default method to compute $A\_p_p\_s_s$. For this aim, the labels $label\_t_i\in\{1,2,3,4,5\}$ are grouped according to their value: in this way, we find the number of voters (tweets) $n\_voters$ for each agreement/disagreement level $A\_level$. $A\_p_p\_s_s$ is the $A\_level$ having the majority of voters; however, if there are more than two $A\_level$s having the same and maximum number of voters, then $A\_p_p\_s_s$ is the nearest integer to the average of all labels $label\_t_i$ (like in \textit{Algorithm 2}).
    where $\Bigl\lfloor ...\Bigr\rceil$ is the round function.
    The majority voting (case \ref{eq:sub_case_majority_voting}) may have a bigger contribution in assigning correct labels than plain standard mean (case \ref{eq:sub_case_mean} taken from Eq. \ref{eq:avg_a_i_labels} of $Alg2$), since it better accounts for class imbalance.
    \item [Algorithm 4] \label{item:alg_4}\textbf{[Alg4]} The previous algorithms take into consideration the neutral label $nl=3$ (\textit{neither disagree, nor agree}) also when $\mid I^p_{tp_s} \mid \neq 0$. However, we wondered how the results would change if $nl$ was \textit{only} considered when $\mid I^p_{tp_s} \mid = 0$. The neutral label may also be assigned in presence of a low number of in-topic $I^p_{tp_s}$: in this particular situation, the user may have not taken a position about the current sentence $s$ yet; also, choosing $A^p_s$ looking at just one tweet may not be significant. Therefore, $Alg4$ stems from $Alg3$ having %with the only adjustment that $A^p_s=nl$ \textit{only} if the number of in-topic tweets is lower than a $min\_num\_tweets$ (value to be defined). 
    \begin{equation}
        l_i = M2(s_i)
    \end{equation}
    where $l_i\in \{1,2,3,4\}$; we define
    %$V_l$ is still computed through Eq. \ref{eq:num_voters}
    \begin{equation}
        a^p_s = 
        \begin{cases}
            3 & \text{if $\mid I^p_{tp_s} \mid < m$}\\ 
            \text{majority voting (case \ref{eq:sub_case_majority_voting})}\\
            \text{rounded standard mean (case \ref{eq:sub_case_mean})}\\
        \end{cases}
    \end{equation}
    where $m$ is the minimum number of tweets for which the majority voting algorithm or the standard mean are executed. Since the $\{3,4\}$ labels in output from $M2(s)$ represent the $agree$ and $completely$ $agree$ final labels, they must be mapped again to the real final integer labels $4$ and $5$ respectively (as coded in Table \ref{tab:agreement_disagreement_labels})
    %Since the algorithm must output $A^p_s=3$ only when $\mid I^p_{tp_s} \mid < m$, $A^p_s$ is set as
    
    \begin{equation}
        A^p_s = 
        \begin{cases}
            a^p_s & \text{if $a^p_s=1\lor a^p_s=2$}\\
            a^p_s + 1 & \text{if $a^p_s=3\lor a^p_s=4$}\\
        \end{cases}
    \end{equation}
    
    %sm & \text{if $sm=1\lor sm=2$}\\
    %sm + 1 & \text{if }\\
    %Finally, $A^p_s$ is
    %\begin{subnumcases}{A^p_s = }
    %    a^p_s + 1 & \text{if $($\ref{eq:sub_case_majority_voting} or \ref{eq:sub_case_mean} were applied$)$ $\wedge$ $a^p_s\in\{3,4\}$}\label{eq:sub_case_redefine_label}\\
    %    a^p_s & \text{otherwise}
    %\end{subnumcases}
    %Case \ref{eq:sub_case_redefine_label} 
    %to adjust the output so that $A^p_s=3$ only when $\mid I^p_{tp_s} \mid < m$.
\end{description}

\section{Experimental Setup}
\label{sec:experimental_setup}
This section first defines the baseline methods to which compare our $T2S$ method. Then, the experiments are described adhering to a set of research questions. Finally, the chosen evaluation metrics are explained.

\subsection{Baselines}
\label{subsec:baselines_description}
It is a good practice to compare the proposed methods with a bunch of baselines. To the best of our knowledge, no baseline method has been devised for the typology of our stance detection task yet: unlike our approach, the state-of-the-art unsupervised user-stance detection method proposed by Darwish et al. \cite{darwish2020unsupervised}  cannot operate without context information from other users and it is not suitable for a multi-class ordinal classification like our case.
%To the best of our knowledge, considering the novelty of the task of inferring the agreement/disagreement level of political sentences from tweets, no baseline method has been devised from the scientific community yet. 
Therefore, the following baselines to compute $A^p_s$ for Party $p$ and sentence $s$ were used:

\begin{description}
    \item [\textbf{Random}] $A^p_s$ is set to a random integer picked from a discrete uniform distribution of \(int\in[1,5]\). The \textit{numpy} random method\footnote{\url{https://numpy.org/doc/stable/reference/random/generated/numpy.random.randint.html}} was used with random seed set to 42.
    %It assigns a random final label to each \((Party, sentence)\) couple. The random integer is picked from a discrete uniform distribution of integer in the interval \([1,5]\). The \textit{numpy} random method was used\footnote{\url{https://numpy.org/doc/stable/reference/random/generated/numpy.random.randint.html}} with random seed set to 42. Besides, each evaluation metric was computed N times, and the standard mean over the N values was considered as the final metric value. Heuristically, N was set to 1000 to be statistically significant.
    \item [\textbf{Predict 3}] $A^p_s$ is set to 3 (\textit{neither disagree, nor agree}).
    %It always assigns the value 3 (\textit{neither disagree, nor agree}) to each (\(Party\), \(sentence\)) couple. %Being the neutral label '3' the middle value of the possible final labels (\(\{1,2,3,4,5\}\)), the Mean Absolute Error (MAE) won't be higher than 2 (see subsection \ref{subsubsubsec:evaluation_metrics} to know the used evaluation metrics): this is a good comparison indicator for our proposed methods.
    \item [\textbf{Sentence Bert}]
    %First, it computes the numeric representation (embedding) of each tweet and each of the 20 sentences. Then, it calculates the cosine similarity between each tweet and sentence; to work on \([0-1]\) values, all cosine similarities are \textit{Min Max Scaled}\footnote{\url{https://scikit-learn.org/stable/modules/generated/sklearn.preprocessing.MinMaxScaler.html}}. Afterwards, for each couple (\(Party_p\), \(sentence_j\)), it sorts the \(Party_p\) tweets by decreasing min-max-scaled value with \(sentence_j\), and it computes the standard mean with the \(top K\) scores. Heuristically, \(K\) was set to 10. Lastly, it maps the obtained value onto the final label:
    %\begin{itemize}
    %    \item label \(1\) (completely disagree) \(if\) \(mean\_value < 0.2\)
    %    \item label \(2\) (disagree) \(if\) \(0.2<= mean\_value < 0.4\)
    %    \item label \(3\) (neither disagree, nor agree) \(if\) \(0.4<= mean\_value < 0.6\)
%        \item label \(4\) (agree) \(if\) \(0.6<= mean\_value < 0.8\)
 %       \item label \(5\) (completely agree) \(if\) \(0.8<= mean\_value <=1\)
  %  \end{itemize}
    The newest Transformer-based language models like BERT can be used as feature extractors \cite{reimers-gurevych-2019-sentence-bert}, providing contextual word and sentence embeddings. %; differently from word representation tools like Bag-of-Words + TF-IDF, Word2Vec\cite{mikolov_2013_word2vec} or GloVe\cite{pennington-etal-2014-glove}, they provide word/sentence contextual embeddings: in fact, thanks to the Attention mechanism \cite{vaswani2017attention}, the numerical representation of each word depends not only by the word itself, but also by the context in which the word occurs. 
    The Sentence-Bert architecture of the \textit{Sentence Transformers} Python library\footnote{\url{https://www.sbert.net/}} was used with the English \textit{all-mpnet-base-v2} model on translated tweets, and with the multi-lingual model \textit{distiluse-base-multilingual-cased-v1} on the Italian tweets. 
    
    Let's define $e_i$ as the embeddings of tweet $t_i\in T_p$ and $e_s$ as the embeddings of sentence $s$. $S$ is the set of cosine similarities between each $e_i$ and $e_s$
    \begin{equation}
        S = \{s_i | s_i = cos\_sim(e_i, e_s)\}^{|T_p|}_{i=1}
    \end{equation}
    Since $s_i\in [-1,1]$, each $s_i$ is $MinMaxScaled$\footnote{\url{https://scikit-learn.org/stable/modules/generated/sklearn.preprocessing.MinMaxScaler.html}} to $ns_i\in[0,1]$. % as
    %\begin{equation}
    %    ns_i = \frac{s_i - min(S)}{max(S)-min(S)}
    %\end{equation}
    Finally, $A^p_s$ is defined as the mapping $M1$ over the average of the $top_K$ scores $ns_i$
    %as $A^p_s = M1(avg)$; $avg$ is computed over the $top_K$ scores $ns_i\in NS$ 
    \begin{equation}
        A^p_s = M1(\frac{\sum^{K}_{i=1}ns_i}{K})
    \end{equation}
    %where
    %\begin{equation}
    %    NS = \{ns_i | ns_1\ge ns_2 \ge ... \ge ns_{|T_p|}\}
    %\end{equation}
    Heuristically, \(K\) was set to 10. 
    %This baseline first computes the embeddings $emb\_t_i$ of each tweet $t_i\in tweets\_p_p$ along with the embeddings $emb\_s_s$ of the current sentence. Then, it calculates the cosine similarity $cos\_sim\_t_i\_s_s$ between each $emb\_t_i$ and $emb\_s_s$; to work on \([0-1]\) values, it \textit{Min Max Scales}\footnote{\url{https://scikit-learn.org/stable/modules/generated/sklearn.preprocessing.MinMaxScaler.html}} all cosine similarity scores ($norm\_cos\_sim\_t_i\_s_s$). Afterwards, it sorts all $tweets\_p_p$ by decreasing min-max-scaled value $norm\_cos\_sim\_t_i\_s_s$, and it computes the standard mean with the \(top_K\) scores ($avg\_top_K$). Heuristically, \(K\) was set to 10. Lastly, $A\_p_p\_s_s$ is the $avg\_top_K$ mapped onto the final label $A\_level$ by calling the \textit{mapping} procedure \ref{alg:mapping_func} with $consider\_label\_3=True$.
    
    %exploits the numeric representation $emb\_t_i$ of each tweet $t_i\in tweets\_p_p$, and the one of the current sentence ($emb\_s_s$) to compute the $A\_p_p\_s_s$ label; 
    % tweet in english/italian -> modelli usati
\end{description}

\subsection{Experiments in detail}
\label{subsec:experiments_detail}
As already explained in section \ref{sec:methodology}, our $T2S$ method has got four parameters to tune: the language model $LM$ to be used for zero-shot classification, the dataset $Di$ from which extract the Twitter timeline $TL_p$, the algorithm $Alg$ for the $Agreement$ step, and the threshold value $th$ for the $Topic$ $Filtering$ step. Considering the values of those parameters in Fig. \ref{fig:Fig2}, we carried out each experiment having in mind the four research questions summarized in Table \ref{tab:experiments_description} and ordered by specificity. 

\begin{table}[!ht]
    \sf\centering
    \caption{Description of all carried out experiments}
    \begin{tabular}{lp{0.6\columnwidth}}
        \toprule
        Experiment Name & Research Question\\
        \midrule
        %\(\langle zsc\_model \rangle\) 
		Best language model $LM$ & Which is the best language model $LM$ for zero-shot classification? Which is the best model to deal with Italian tweets? All in all, is an English model better?\\
		Best dataset $D$ & Fixed the language model $LM$, which is the best dataset to work on, considering all proposed algorithms? Hence, which is the best time period to listen to before a Political Election?\\
		Best algorithm $Alg$ & Fixed the language model $LM$ and dataset $Di$, which is the best algorithm to work on, considering all evaluated thresholds $th$? Are all our proposed algorithms better than the baselines (subsection \ref{subsec:baselines_description})? Are the more complex algorithms better or not?\\
		Best threshold $th$ & Fixed the language model $LM$, the dataset $D-Nmonths$ and the algorithm $Alg$, which is the best filtering threshold $th$, hence the optimal set-up? \\
		%Party and Sentence Analysis & Using the best set-up, analyze the fine-grained results over each Party and Sentence.\\
		Party and Sentence Analysis & Fixed the optimal setup for our framework, which are the Parties on which $T2S$ behaves well or poorly? What about the sentences?\\
		\bottomrule
    \end{tabular}
    \label{tab:experiments_description}
\end{table}

\subsection{Evaluation}
Generally, the performance of a prediction model is evaluated through several error metrics such as the Mean Squared Error (MSE), the R2 Score, Residual Plots and/or the Mean Absolute Error (MAE); moreover, \cite{baccianella_2009_macro_avg_mae} suggests to use the \textit{Macro Averaged MAE} for Ordinal Regression tasks with imbalanced classes. 

%Ordinal Regression = Ordinal Classification
Our task of predicting the agreement/disagreement level is close to Ordinal Regression because there's an inherent ordering between classes (\{\textit{completely disagree, disagree, neither disagree nor agree, agree, completely agree}\}); however, it differs from an ordinal regression because there are meaningful differences between classes: for example, predicting \textit{disagree} instead of \textit{completely disagree} is a more acceptable error than predicting \textit{neither disagree nor agree} or \textit{agree}/\textit{completely agree}. Besides, the class imbalance is not an issue for our task: it doesn't matter if more \textit{disagree} or \textit{agree} labels (classes) $G^p_s$ must be predicted, the point is measuring the error between the predicted ($A^p_s$) and true ($G^p_s$) values as fair as possible. For these reasons, the Macro Averaged MAE was not used. Also, the other error metrics do not fit well our task; however, since an appropriate error metric has not been devised yet, the \textit{MAE} metric was employed. The MSE was used at first too, but it was just a replica of the MAE result, so it was discarded.

%\begin{description}
%    \item [\textit{Mean Absolute Error}] is the expected value of the absolute error loss. If \(\hat{y}_i\) is the predicted value of the \(i^{th}\) sample, and \(y_i\) is the corresponding true value, then the MAE estimated over \(n_{samples}\) is defined as:
%    \begin{equation}
%        MAE(y,\hat{y}) = \frac{1}{n_{samples}}\cdot\sum_{i=0}^{n_{samples}-1} |y_i-\hat{y}_i|
%        \label{eq:mae}
%    \end{equation}
%    Low values of MAE \(\in[0-1]\) indicate good performances.
%\end{description}
%baccianella_2009_macro_avg_mae, willmott_2005_mae
% una metrica più vicina al nostro task sarebbe stata meglio... però abbiamo pensato di attenerci alle metriche solitamente usate.

Lastly, since the predicted value is an integer among \(\{1,2,3,4,5\}\), a classification evaluation metric was considered as well: the weighted \textit{F1} score was picked, since it summarizes both Precision and Recall \cite{sebastiani2002machine}. % descrizione di F1 weighted
%\begin{description}
%    \item [\textit{F1\_score}] can be interpreted as a harmonic mean of the precision and recall. The relative contribution of precision and recall are equal. For our multi-class case, the final F1\_score is the average of the F1\_score of each class weighted by support (the number of true instances for each class). This choice accounts for label imbalance: for the F1\_score, the class imbalance must be taken into consideration.
%    \begin{equation}
%        F1\_class_c = 2\cdot\frac{precision(y,\hat{y})\cdot recall(y,\hat{y})}{precision(y,\hat{y})+recall(y,\hat{y})}
%        \label{eq:f1_class_score}
%    \end{equation}
%    \begin{equation}
%        F1\_weighted = \frac{\sum_{c=1}^{n_{classes}} F1\_class_c \cdot num\_true\_instances\_class_c}{\sum_{c=1}^{n_{classes}} num\_true\_instances\_class_c}
%        \label{eq:f1_weighted_score}
%    \end{equation}
%    Low values of F1 \(\in[0-1]\) indicate good performances.
%\end{description}

The \textit{sklearn.metrics} Python package was used to compute both \textit{MAE}\footnote{\url{https://scikit-learn.org/stable/modules/generated/sklearn.metrics.mean\_absolute\_error.html}} 
and \textit{F1\_weighted}\footnote{\url{https://scikit-learn.org/stable/modules/generated/sklearn.metrics.f1\_score.html}}.

\section{Results and Discussion}
\label{sec:results_and_discussion}
The results are discussed adhering to the experiments summarized in Table \ref{tab:experiments_description}. We also carried out the same experiments considering different tweet typologies (e.g., without retweets, which represented a large percentage of the total tweets), but we obtained worse results than using all of them (original tweets, retweets, replies, quotes).

\subsection{Best Language Model LM}
\begin{figure}[!ht]
    \centering
    \begin{subfigure}[c]{\columnwidth}
        \centering
        \includegraphics[scale=0.5]{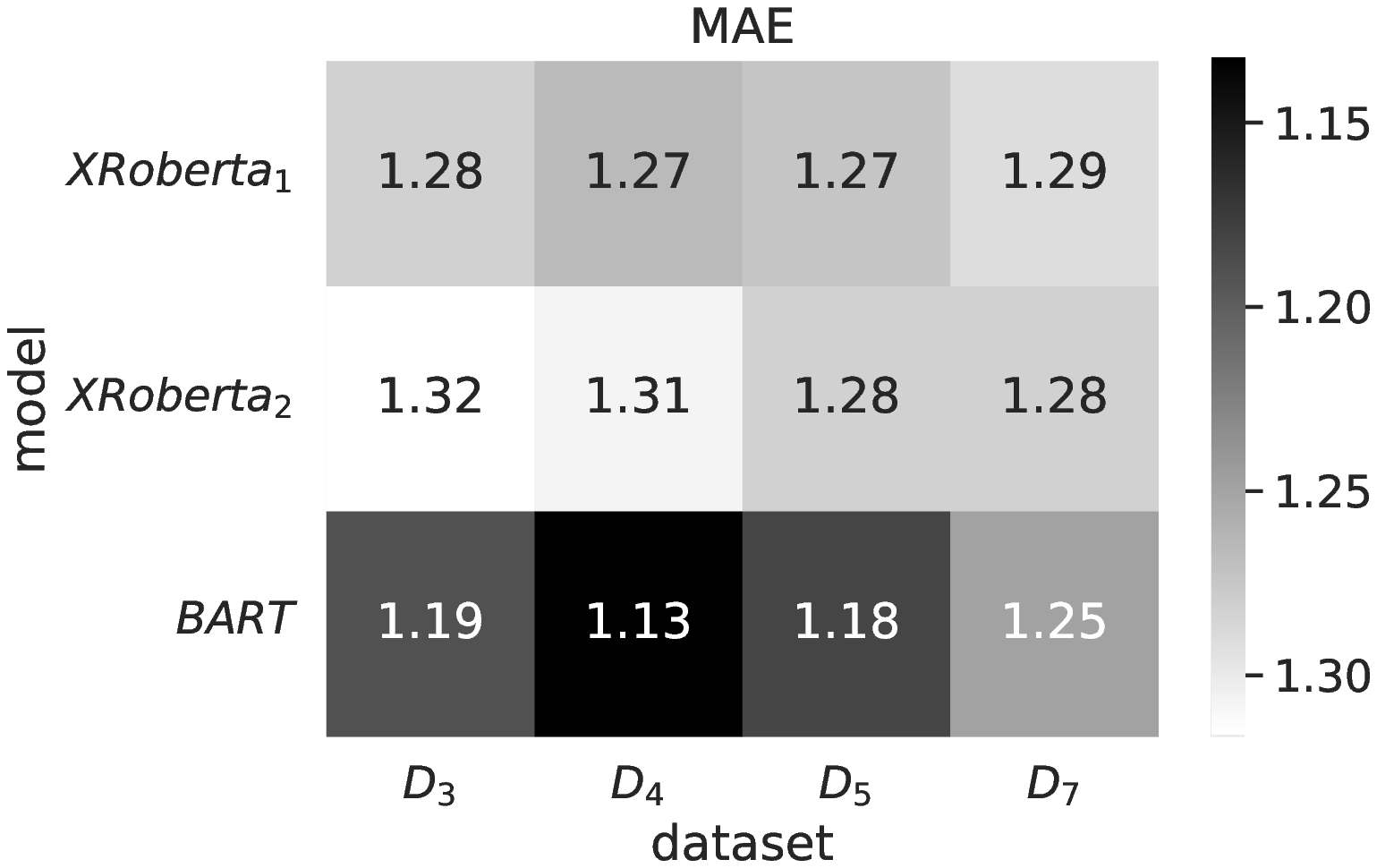}
    \end{subfigure}
    \begin{subfigure}[c]{\columnwidth}
        \centering
        \includegraphics[scale=0.5]{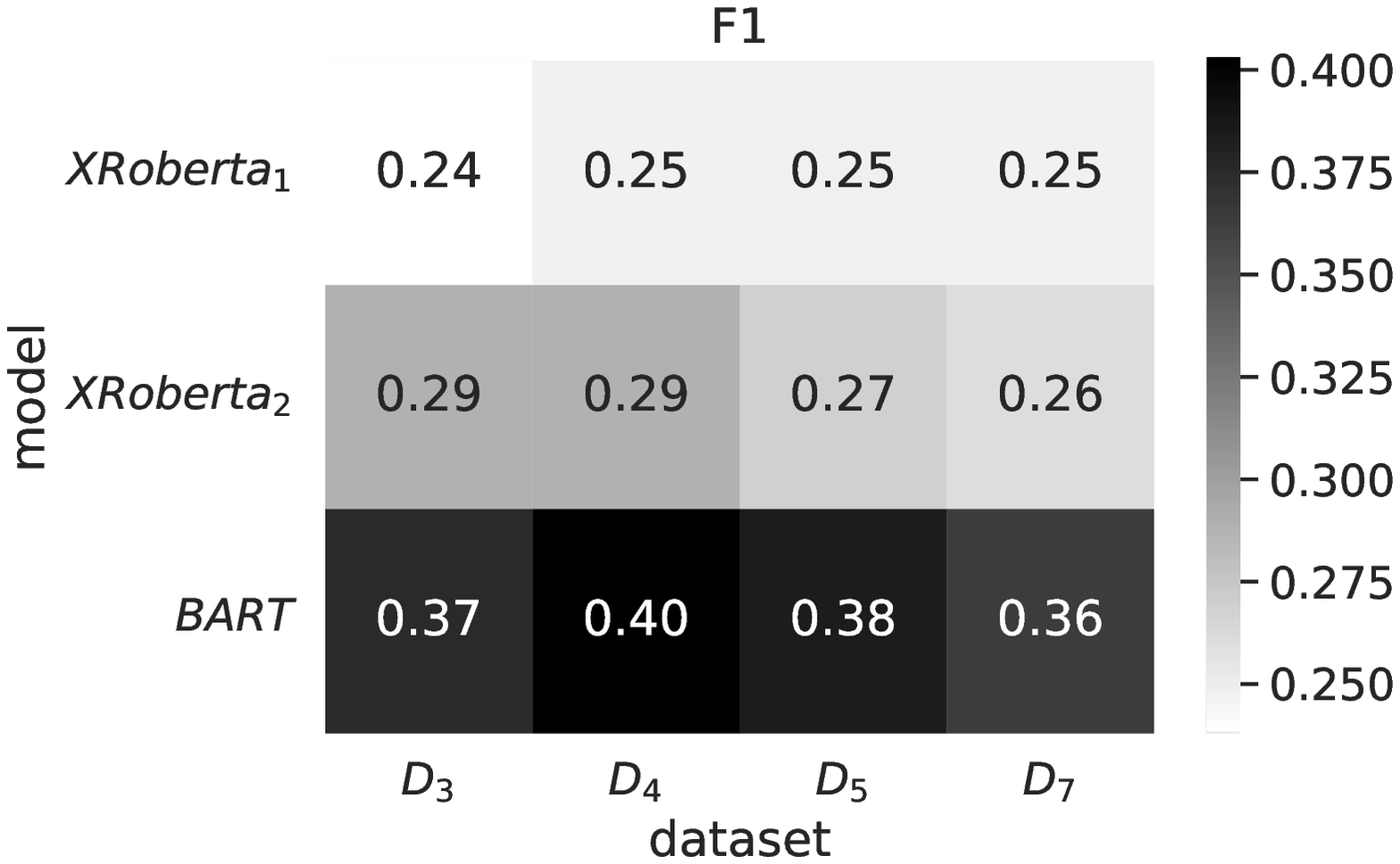}
    \end{subfigure}
    \caption{Best MAE and F1 values of our $T2S$ method for each couple $(LM_i,D_j)$ of language models and datasets. Darker colors indicate optimal values for both metrics.}
    %for Best language model LM, Best dataset D
    \label{fig:Fig4}
\end{figure}

First, we asked ourselves which is the best language model to be used for zero-shot classification: among cross-lingual language models employed on Italian tweets, is it better a model pre-trained on a mixture of languages including Italian, or a model further fine-tuned on an Italian corpus? Furthermore, would the results benefit from using an English language model on translated tweets instead? We answered these questions by looking at Fig. \ref{fig:Fig4}: each cell $(LM_i,D_j)$ indicates the minimum MAE (maximum F1) obtained with our $T2S$ method for a certain language model $LM_i$ and dataset $D_j$ by varying the algorithm $Alg$ and the threshold $th$ according to Fig. \ref{fig:Fig2}.

Among the cross-lingual models $XRoberta_1$ and $XRoberta_2$, the best one seemed to be $XRoberta_1$: it had an overall better MAE, while F1 results were close to $XRoberta_2$'s; we considered MAE as the first metric to judge the performances, since it tells how much we are close to the correct answer. Apparently, fine-tuning on an Italian translation of a subset of the MNLI dataset ($XRoberta_2$) doesn't contribute a lot to text classification in our $T2S$ framework.
%MAE it's the main metric for prediction tasks like ours

All in all, the best choice is translating the pre-processed tweets in English and using an English model like $BART$: it reached significantly higher values on both MAE and F1. Supposedly, using a model pre-trained and fine-tuned on a single language gives better results for our prediction task: learning on a single language allows to focus on more details, features of the language.

\subsection{Best Dataset D}
% referring to Fig. \ref{fig:Fig5_MAE} and \ref{fig:Fig5_F1}
Fixed the language model $LM=BART$, the dataset $D_4$ was immediately detected as the best one, since it had the best MAE and F1 (Fig. \ref{fig:Fig4}). Presumably, the Twitter political discussion since four months before the Italian elections were enough to grasp the Parties' stances.

We evaluated the mean MAE and mean F1 for each cell $(LM_i,D_j)$ of Fig. \ref{fig:Fig4} as well, but the results confirmed $BART$ and $D_4$ as the best language model and dataset.

\subsection{Best Algorithm Alg}
\begin{figure}[!ht]
    \centering
    \begin{subfigure}[c]{\columnwidth}
        \centering
        \includegraphics[scale=0.5]{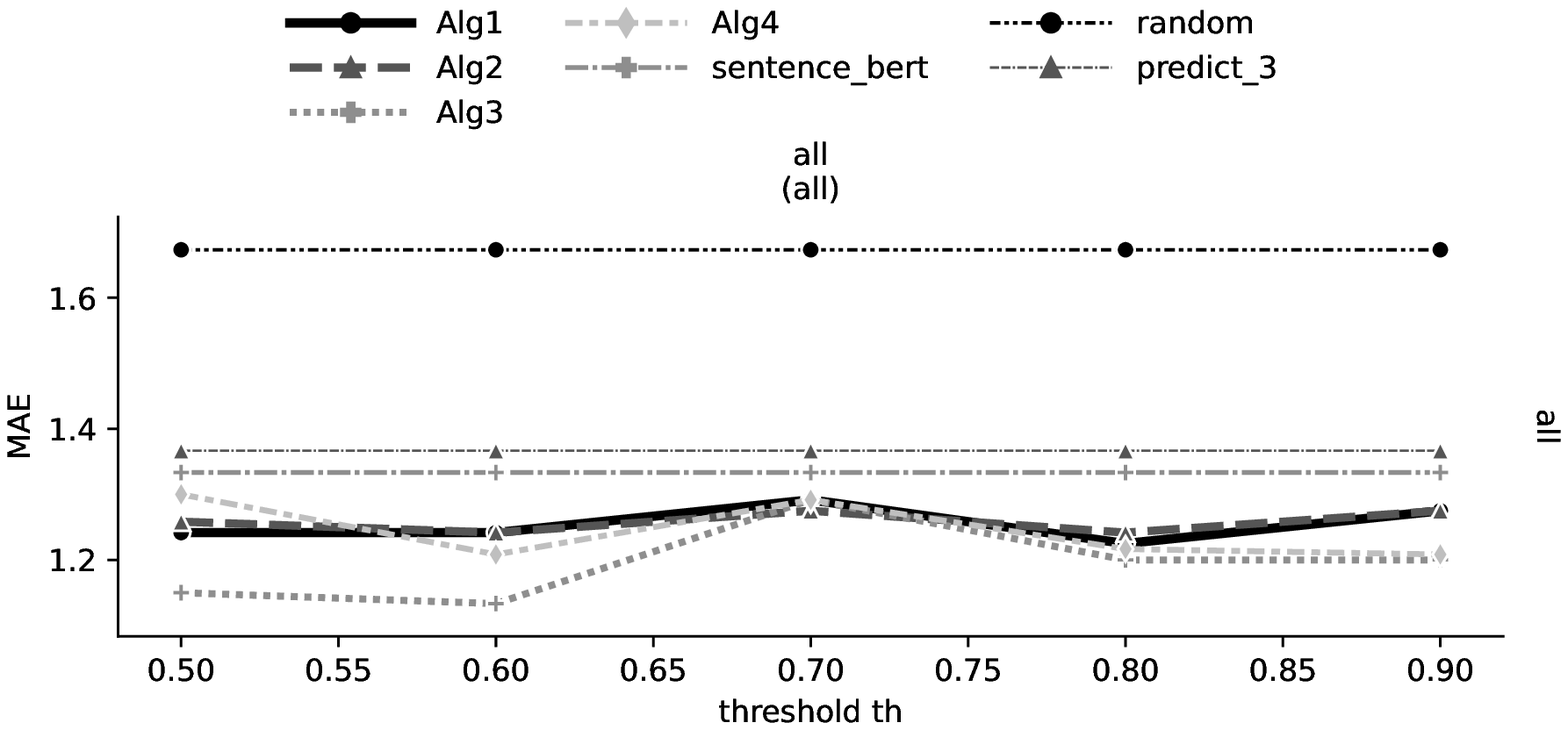}
    \end{subfigure}
    \begin{subfigure}[c]{\columnwidth}
        \centering
        \includegraphics[scale=0.5]{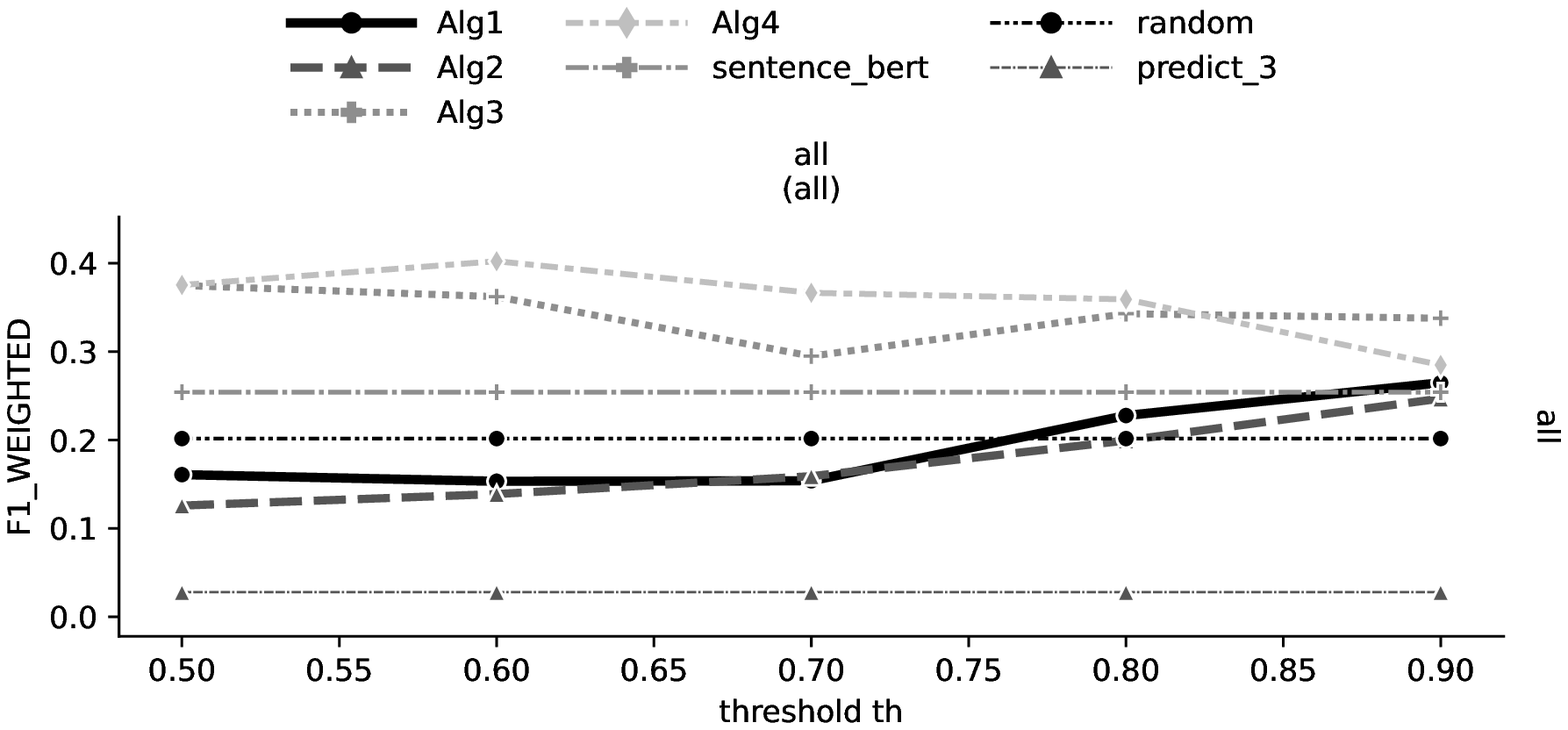}
    \end{subfigure}
    \caption{MAE and F1 of our four proposed algorithms $Alg$s and the three baselines by varying the threshold $th$. It is shown $Alg4$ with $m=3$.}
    %for Best language model LM, Best dataset D
    \label{fig:Fig5}
\end{figure}

Once the language model and dataset were chosen, we wondered whether our algorithms $Alg$s are better than the three baselines $random$, $predict\_3$ and $sentence\_bert$, and which is the best $Alg$ considering all thresholds $th$. To this aim, we fixed the language model $LM=BART$ and dataset $D_4$, and we varied the algorithms $Alg$ and thresholds $th$s; Fig. \ref{fig:Fig5} describes how much each algorithm performed across different thresholds. %Each point indicates
%\begin{equation}
%    metric(F_{su}, G) |_{su=\{BART, D_4, Alg, th\}}
%\end{equation}
%where $metric\in\{MAE, F1\}$ and $Alg$ and $th$ vary. 
These results include the performances of the three baselines as well.

As far as it concerns the MAE, all proposed $Alg$ performed much better than the three baselines. However, only $Alg3$ and $Alg4$ surpassed the baselines on F1, indicating that aggregating the contribution of each tweet ($l_i$) provides a better prediction precision than directly weight-averaging on the sentence scores $s_i$ of each tweet. %indicating that micro-aggregating the contribution of each tweet ($al_i$) provides a better prediction precision than macro-aggregating on the sentence scores $ss_i$ of each tweet. 
Altogether, the optimal algorithm can be identified in $Alg3$: F1 seemed to contradict it and bend over $Alg4$ instead, but the gain over the prediction error is far more important. This result suggests that assigning the neutral label (\textit{neither disagree, nor agree}) only when there's a minimum number of tweets $m$ does not boost the performance of our $T2S$ method. Also, we executed $Alg4$ with $m=\{2,3\}$, finding out that the results didn't vary a lot from each other; therefore, we showed $Alg4_{m=3}$ in Fig. \ref{fig:Fig5}.

\subsection{Best Threshold th}
Fixed the language model $LM=BART$, the dataset $D_4$ and the algorithm $Alg3$, threshold $th=0.6$ was immediately detected as the optimal one, since it had the best MAE and a good F1 (Fig. \ref{fig:Fig5}).

Therefore, the best setup $su_{opt}$ of our $T2S$ framework was $(LM,$ $D_j,$ $Alg,$ $th)=(BART,$ $D_4,$ $Alg3,$ $0.6)$

\subsection{Party and Sentence Analysis}
\subsubsection{Party}
\begin{figure}[!ht]
    \centering
    \begin{subfigure}[c]{\columnwidth}
        \centering
        \includegraphics[scale=0.5]{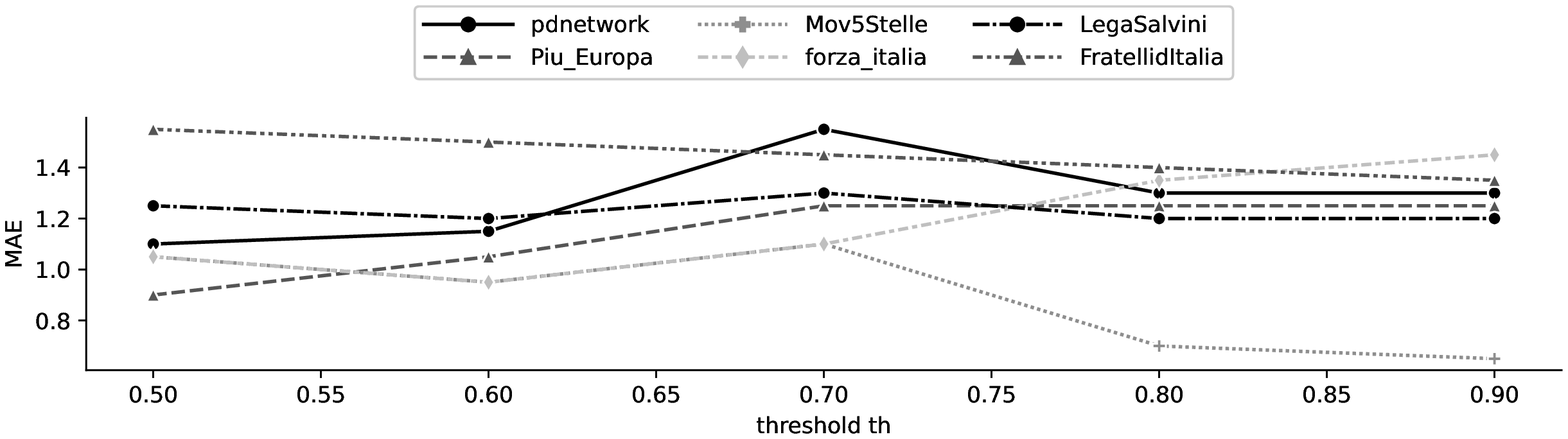}
    \end{subfigure}
    \begin{subfigure}[c]{\columnwidth}
        \centering
        \includegraphics[scale=0.5]{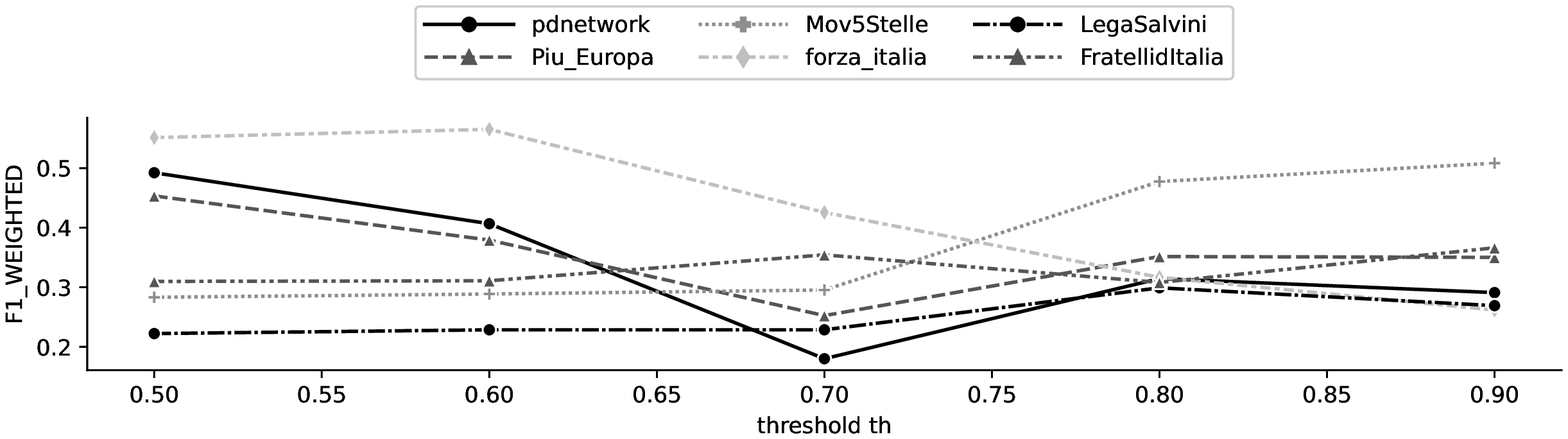}
    \end{subfigure}
    \caption{MAE and F1 computed for each Party over the stance predictions of the $20$ VAA statements. The optimal $su_{opt}$ is used, but the threshold $th$ varies.}
    %MAE and F1 computed for each Party over the $F^p_{su}$ predictions. The optimal $su_{opt}$ is used, but the threshold $th$ varies
    %for Best language model LM, Best dataset D
    \label{fig:Fig6}
\end{figure}

To explore the specific performance of our $T2S$ method over the Parties, we used the optimal setup $su_{opt}$ but by varying the threshold $th$. Fig. \ref{fig:Fig6} shows the results. Each point indicates the MAE (F1) on the 20 sentences' agreement level $A^p_s$ for a certain Party $p$.

Each Party behaves differently, thus it is likely that $T2S$ highly depends on the Party's timeline in terms of how much it generally writes, how much it writes in-topic, how much it writes using figures of speech or hashtags and emojis (which we removed). Looking at both the MAE and F1, we observed a regular trend for thresholds $th=\{0.8,0.9\}$ for five parties out of six: the outlier Party $Mov5Stelle$ was more predictable for those thresholds. That may happen because the user's timeline deal with a certain statement in a clearer way; for example, looking at $Mov5Stelle$ and $forza\_italia$'s tweets filtered for the sentence $S19$ and $th=0.9$, we saw that $Mov5Stelle$ wrote clearer and explicit tweets supporting the argument (it completely agrees), while from $forza\_italia$'s timeline it's not immediately clear that it disagrees; $forza\_italia$ tweeted about tax reduction, fewer fees on families, and job creation, but it didn't explicitly say that it disagrees with income support for the poorest sections of the population being a positive trait for the Italian economy, thus our $T2S$ framework assigned it 'completely agrees'.
% 

%; each point indicates
%\begin{equation}
%    metric(F^p_{su}, G) |_{su=\{BART, D_4, Alg3, th\}}  
%\end{equation}
%where $F^p_{su}$ is the \(feature\_array\) of $|S=\{$ sentences defined in \ref{subsec:topic_filtering}$\}|$ elements of Party $p$. $F^p_{su}$'s elements are triplets $(p,s,A^p_s)$ for set-up $su$, where the sentence $s$ varies.

\subsubsection{Sentence}
\begin{figure}[!ht]
    \centering
    \includegraphics[scale=0.5]{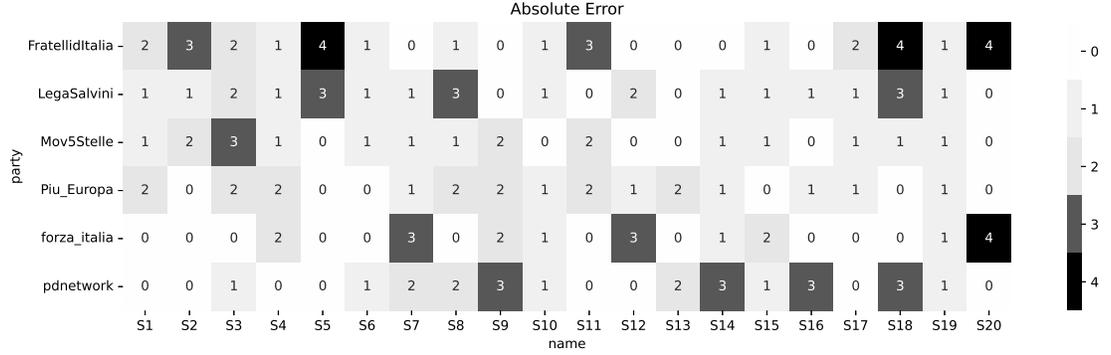}
    \caption{Absolute Error for each (party, sentence) couple using the optimal setup $su_{opt}$.}
    \label{fig:Fig7}
\end{figure}

\begin{figure}[t]
    \centering
    \begin{subfigure}[c]{0.3\linewidth}
        \centering
        \includegraphics[scale=0.3]{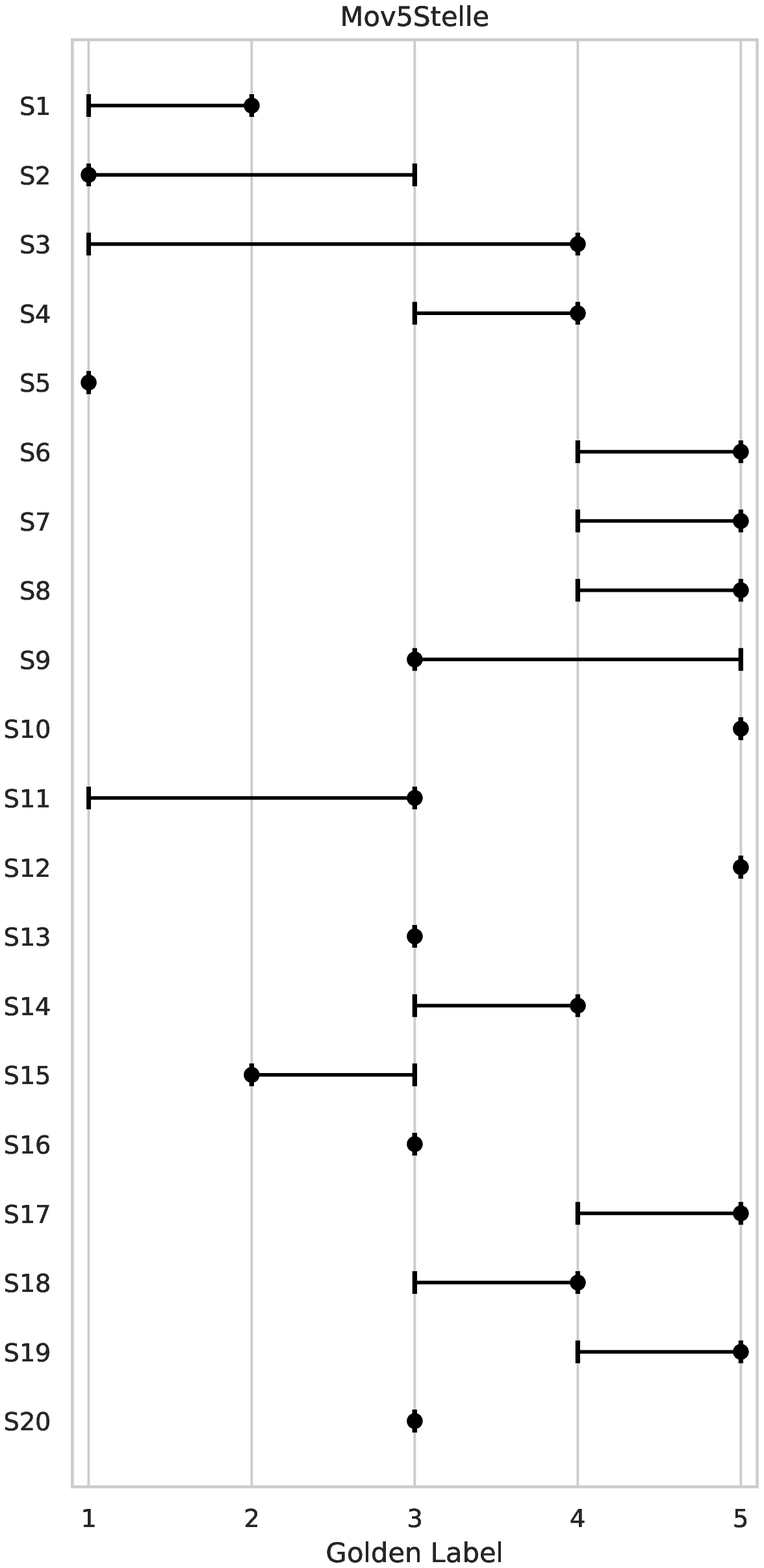}
    \end{subfigure}
    \hfil
    \centering
    \begin{subfigure}[c]{0.3\linewidth}
        \centering
        \includegraphics[scale=0.3]{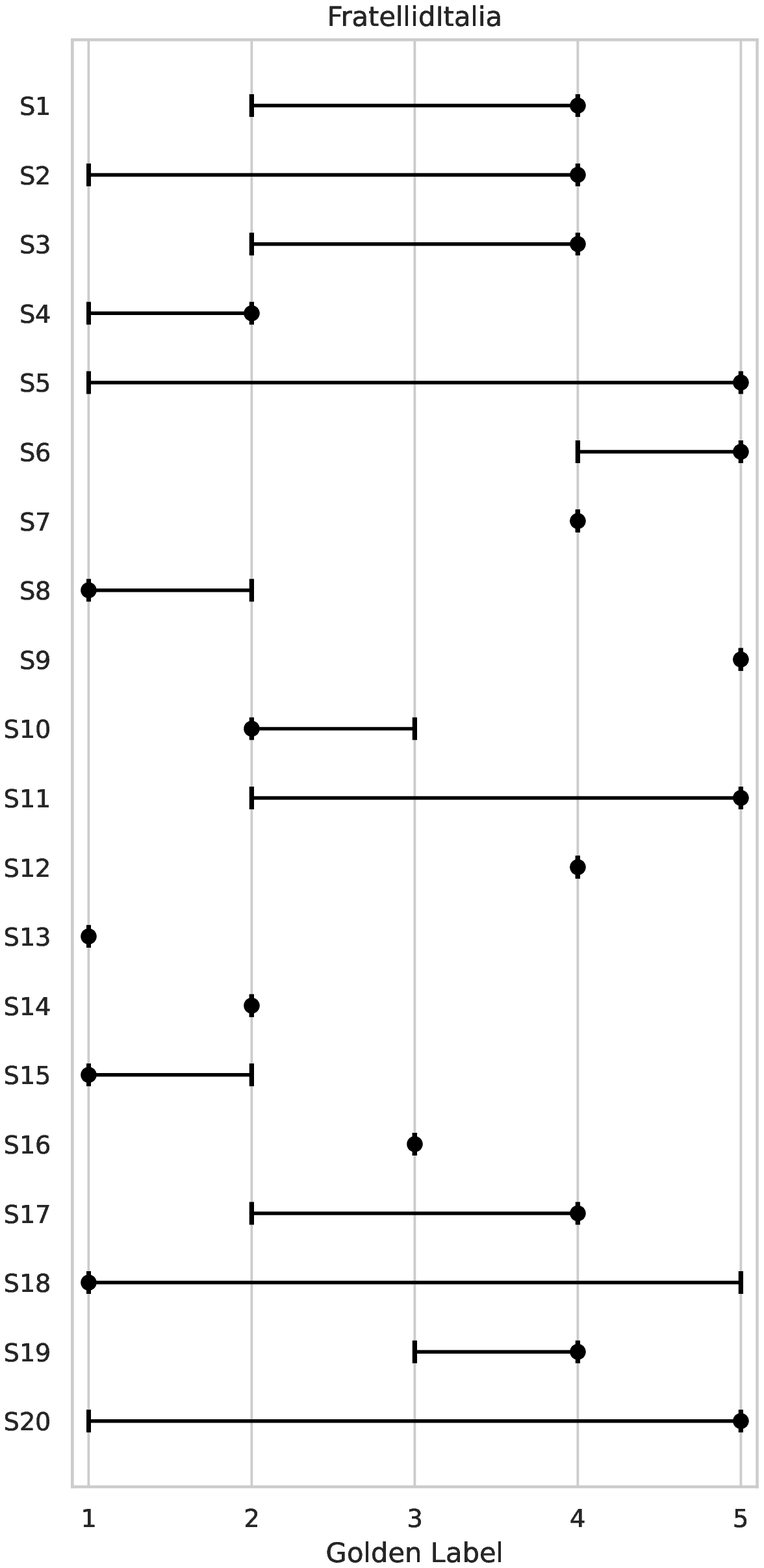}
    \end{subfigure}
    \hfil
    \centering
    \begin{subfigure}[c]{0.3\linewidth}
        \centering
        \includegraphics[scale=0.3]{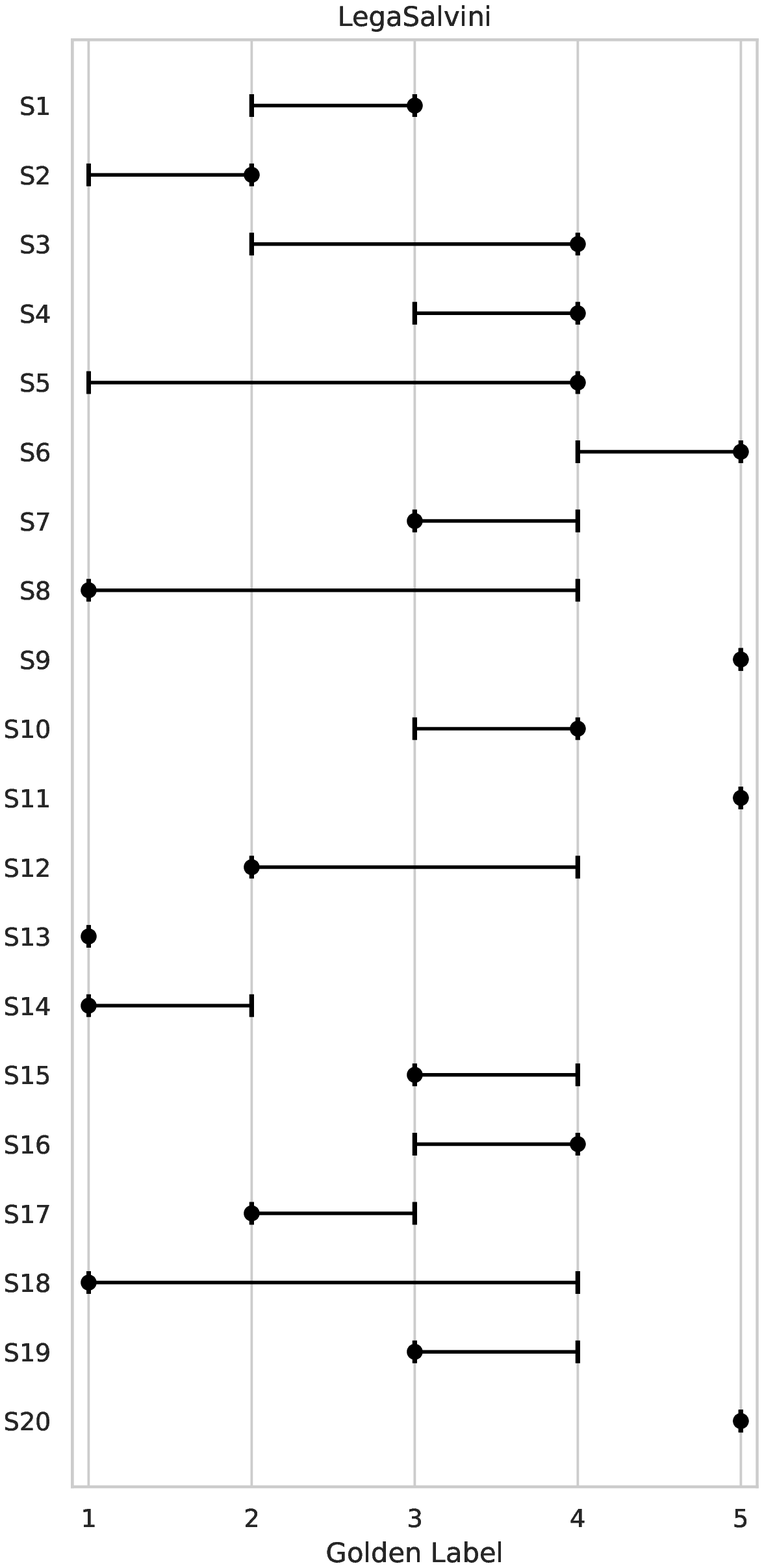}
    \end{subfigure}
    \caption{Error bars for $Mov5Stelle$, $FratellidItalia$ and $LegaSalvini$ using $su_{opt}$: $\blacklozenge$ indicates the ground-of-truth for a sentence, while $|$ the predicted label.}
    \label{fig:Fig8}
\end{figure}

To explore the specific performance of our $T2S$ method over the sentences and Parties, we used the optimal setup $su_{opt}$: Fig. \ref{fig:Fig7} presents the absolute error for each (Party, sentence), while Fig. \ref{fig:Fig8} shows the error bars for the Parties on which $T2S$ performed better ($Mov5Stelle$) and worse ($FratellidItalia$), and the Party having the highest number of absolute errors equal to '1' ($LegaSalvini$); Fig. \ref{fig:Fig10} displays the number of in-topic tweets in output from the $Topic$ $Filtering$ step. 

The acceptable error value is '1', since on a five-valued Likert scale (Table \ref{tab:agreement_disagreement_labels}) being wrong by one level does not have a great impact on the agree/disagree levels, unless the correct label is the intermediate (neutral) one. To understand whether the '1' errors are acceptable or not, an error measure taking into account that predicting \textit{disagree} instead of \textit{completely disagree} is a more acceptable error than predicting \textit{neither disagree nor agree} or \textit{agree}/\textit{completely agree} could have provided more insightful hints; however, such an error measure has not been devised yet in literature. Therefore, such insights can be analyzed by looking at error bars like the ones in Fig. \ref{fig:Fig8}: again, the goodness of the '1' absolute errors depended on the Party's timeline, with $Mov5Stelle$ having $70\%$ of good '1', while $LegaSalvini$ only the $20\%$.

%Mov5Stelle, 10 uni, 3 non buoni (30%)
% FratellidItalia, 6 uni, 2 non buoni (33%)

\begin{figure}[!ht]
    \centering
    \includegraphics[scale=0.5]{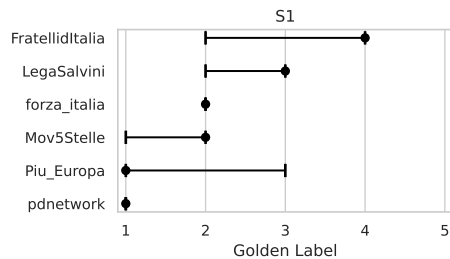}
    \caption{Error bars for $S1$ "overall, being EU members is a disadvantage" (topic: "European Union disadvantages"). $su_{opt}$ has been used. $\blacklozenge$ indicates the ground-of-truth for a sentence, while $|$ the predicted label.}
    \label{fig:Fig9}
\end{figure}

\begin{figure}[!ht]
    \centering
    \includegraphics[scale=0.5]{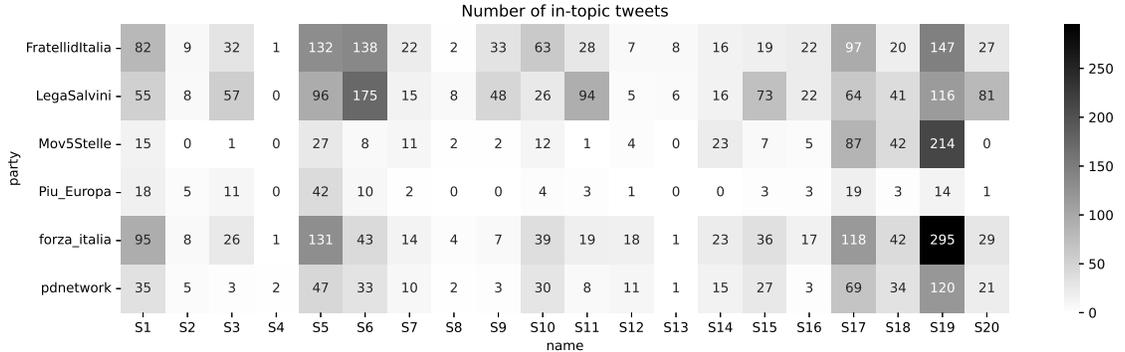}
    \caption{Number of tweets over which each $A^p_s$ is computed. The optimal setup $su_{opt}$ is used.}
    \label{fig:Fig10}
\end{figure}

%S1: overall, being EU members is a disadvantage (chosen this sentence because it's clear that there's no correlation between the absolute error and the number of tweets).
% topic: European Union disadvantages
Looking at both Fig. \ref{fig:Fig7} and \ref{fig:Fig10}, it seems that there's no correlation between the error magnitude and the number of tweets. $S1$ clearly shows this, therefore we investigated how $T2S$ with $su_{opt}$ took the decision about each Party (Fig. \ref{fig:Fig9}): $pdnetwork$'s prediction 'completely disagree' is correct, indeed it wrote lot of pro Europe tweets; $forza\_italia$ tweeted about the need to change Europe, hence the 'disagree' label correctly assigned; $piu\_europa$ completely disagrees with $S1$, but it was predicted as 'neither disagree, nor agree': out of 18 tweets, five talked about the European countries not having neither common foreign policy nor common defence and were labelled as 'neither disagree, nor agree', eight were pro Europe (five labelled as 'disagree', three labelled as 'completely disagree'), while the remaining ones talked about Europe in a negative way and hence labelled either as 'agree' or 'completely agree'. In this case, the standard mean was applied (see Eq. \ref{eq:sub_case_mean} of $Alg3$) and rounded to '3' ('neither disagree, nor agree'); this case suggests that more tweets supporting the correct stance are needed to correctly predict the $A^p_s$ label. The majority of $Mov5Stelle$'s tweets were pro Europe, hence our $T2S$ predicted 'completely disagree' instead of 'disagree'. Instead, $LegaSalvini$ is neutral about $S1$ but it was predicted as in disagreement: the majority of its tweets talked about the need to change Europe; we believe that it's difficult to predict neutrality. Setting the neutral label only when there are no in-topic tweets surely helps, and $Alg4$ should have brought advantages in this regard; nonetheless, $Alg4$'s $m$ param is not easy to set, since it may depend on how a Twitter user expresses itself. Last but not least, $FratellidItalia$ tweeted a lot about the need to change Europe, hence it was predicted as in disagreement with $S1$ (as correctly done for $forza\_italia$); however, its real stance is 'agree'. $FratellidItalia$'s tweets classified as in agreement with $S1$ have got words like 'failure', disappointment', 'save from', 'penalized', 'disappear' together with 'Europe', 'EU' or 'European'; again, we found another proof regarding the dependency of our $T2S$ framework with how a Twitter user communicates its stance about something.

% A questo punto selezionando i casi di predizione peggiore, vogliamo provare a discutere il risultato analizzando la specifica sentence coinvolta?
%[DA QUALCHE PARTE DIRE CHE ABBIAMO PROVATO ANCHE AD USARE PARTE DELLA TIMELINE (e.g. togliendo i retweet) MA ABBIAMO OTTENUTO RISULTATI PEGGIORI]

\subsection{Possible limitations of our framework}
The usage of the \textit{Tweets2Stance} framework has to be seen in light of some limitations that can be addressed in several ways. First and foremost, the availability of the official position of a social media user $u$ (here an Italian Party) about a social-political statement $s$ is crucial to measure our framework's performance; if we wanted to generalize T2S to social media users different from political parties and dealing with other typologies of statements (e.g., about extremist ideologies), two steps could be followed: a domain expert would be needed to accurately design the statements, and users should voluntarily express their opinions through a VAA-like questionnaire. Second, it is highly probable that using a more accurate translation model than the plain Google Translator would increase the performances. Finally, a ZSC may not be pre-trained on current hot-topics and new coined words, hence it performs poorly on those arguments; we encountered this issue when looking at results for sentence $s13$ (Table \ref{tab:topic_definition_eng}) about the automatic citizenship of children who were born in Italy to foreign parents (related to the Italian decree-law called "DDL Zan"): the three tested ZSC models didn't recognize tweets containing "DDL" and "Zan" words as associated to sentence $s13$. This limitation can be addressed in the future by using the most sophisticated and updated ZSC models, which hopefully will deal with current hot-topics.
%hashtag$ potrebbero essere tradotti in inglese, ma è difficile dall'italiano.
%It is worth noticing that the major issue of our framework are two: the availability of the ground-of-truth for each couple ($u$, $s$) for testing purposes, and the chosen translation model to English texts; about the latter, a multi-language ZSC model can be employed in our framework, but the final performance is lower than using a ZSC model trained on English texts only.
%da inserire: un sistema di topic modeling potrebbe non essere allenato su nuovi argomenti (e.g. ddl zan). Collegato a:
%da inserire: tecnica è sempre la stessa ma può aumentare l'accuratezza mano a mano che escono modelli di ZSL migliori

\section{Conclusions and Future Works}
\label{sec:conclusion}
In this work we introduced Tweets2Stance, a novel unsupervised stance-detection framework which uses the timeline of a Social Media user to infer its stance towards a reference political-social statement on a 5-value Likert scale. In particular, we dealt with the stance of 20 political statements for the six major parties in Italy. Results showed that, although the general maximum F1 value was $0.4$, $T2S$ could correctly predict the stance with a general minimum MAE of $1.13$, which is a great achievement considering that MAE tells how much we are close to the correct answer, and that we worked with a final five-valued label. Also, as we hypothesized, the $T2S$'s performance highly depends on how the Twitter account of the Party (hence the social media user) writes, e.g. the employed figures of speech, the words used, and so on. Notice that we removed both hashtags and emojis because the ZSC component had issues in processing them, since it was not pre-trained on social media texts; however, since they may contain useful information to the ZSC component, it would be worth trying to include and normalize both of them by separating each hashtag's words and removing the $\#$ symbol, and by substituting each emoji with words. 
As mentioned when introducing the work, our major goal herein is to design and evaluate, through multiple experiments, a framework working as a stance detector potentially generalizable to several topics. If applied to political discourse, it could cover the first step of a pipeline whose output is the user's political leaning. In the near future, we will investigate how \textit{T2S}'s agreement levels output can be used to that aim.
%to derive the political leaning of a social media user, for example by trying to emulate a VAA algorithm.
%In the near future, we will investigate the second step emulating a VAA algorithm and focusing over the impact of the stance-detector performance on the global goodness of the complete framework as a political leaning predictor.
Besides, we hope to apply it to detect extremists accounts on social media; however, a domain-expert may be needed to define precise social statements to use.
At last, our $T2S$ coped-with Italian tweets by translating them to English, hence it can be applied to English tweets as well, and hopefully reaching better performances.

%da inserire: Also, the ZSC component may benefit from including the normalized hashtags, i.e. separating each hashtag's words and removing the $\#$ symbol. 

%final five-valued label
%predicting a five-valued stance is a challenging task, reaching a general minimum MAE of $1.13$ and a maximum F1 of $0.4$.

%T2S framework
%T2S framework results on parties
% using these results for political leaning
% use T2S to detect extremists on social media, especially in Italy.

\section*{Acknowledgments}
We are immensely grateful to the Observatory on Political Parties and Representation \cite{political_observatory} for providing us the 2019 official position of the six major Italian parties about 20 political statements \cite{Navigatore_Elettorale_2019}.
%List here those individuals who provided help during the research (e.g., providing language help, writing assistance or proof reading the article, etc.).
%\section*{Funding Sources}

This work was supported by the SoBigData++: European Integrated Infrastructure for Social Mining and Big Data Analytics project [871042].
%Funding: This work was supported by the National Institutes of Health [grant numbers xxxx, yyyy]; the Bill & Melinda Gates Foundation, Seattle, WA [grant number zzzz]; and the United States Institutes of Peace [grant number aaaa].

%\section{Bibliography styles}

%There are various bibliography styles available. You can select the style of your choice in the preamble of this document. These styles are Elsevier %styles based on standard styles like Harvard and Vancouver. Please use Bib\TeX\ to generate your bibliography and include DOIs whenever available.

%Here are two sample references: \cite{Feynman1963118,Dirac1953888}.

%\section*{References}

\bibliography{mybibfile}

\end{document}